\RequirePackage{fix-cm}
\RequirePackage{amsmath}
\documentclass[twocolumn,epjc3]{svjour3}
\smartqed  
\RequirePackage{graphicx}

\usepackage[utf8]{inputenc}
\usepackage{url}
\usepackage{amsmath}
\usepackage{xcolor}
\usepackage{cite}
\usepackage[english]{babel}

\newcommand\sqrts{$\sqrt{s_\mathrm{NN}}$ }
\newcommand\smsh{\texttt{SMASH} }
\newcommand\hybrid{\texttt{SMASH-vHLLE-hybrid} }
\newcommand\vhlle{\texttt{vHLLE} }
\newcommand\sampler{\texttt{SMASH-hadron-sampler } }

\newcommand\dndy{$\mathrm{dN}/\mathrm{d}y$ }
\newcommand\dndmt{$\mathrm{dN}/\mathrm{d}m_\mathrm{T}$ }
\newcommand\mt{$m_\mathrm{T}$ }
\newcommand\pt{$p_\mathrm{T}$ }
\newcommand\meanpt{$\langle p_\mathrm{T} \rangle$ }
\newcommand\midyyield{$\mathrm{dN}/\mathrm{d}y|_{y = 0}$ }

\journalname{Eur. Phys. J. }
\begin{document}

\title{Particle production in a hybrid approach for a beam energy scan of Au+Au/Pb+Pb collisions between \sqrts = 4.3 GeV and \sqrts = 200.0 GeV
}


\author{Anna Sch\"afer\thanksref{e1, addr1, addr2, addr5}
        \and
        Iurii Karpenko\thanksref{addr3} 
        \and
        Xiang-Yu Wu\thanksref{addr4}
        \and
        Jan Hammelmann\thanksref{addr1, addr2}
        \and
        Hannah Elfner\thanksref{addr5, addr1, addr2, addr6}
}

\thankstext{e1}{e-mail: aschaefer@fias.uni-frankfurt.de}

\institute{Frankfurt Institute for Advanced Studies (FIAS), Ruth-Moufang-Stra{\ss}e 1, 60438 Frankfurt am Main, Germany \label{addr1}
           \and
           Institut f\"ur Theoretische Physik, Johann Wolfgang Goethe-Universit\"at, Max-von-Laue-Stra{\ss}e 1, 60438 Frankfurt am Main   \label{addr2}
           \and
           GSI Helmholtzzentrum f\"ur Schwerionenforschung, Planckstrasse 1, 64291 Darmstadt, Germany \label{addr5}
           \and
           Faculty of Nuclear Sciences and Physical Engineering,
           Czech Technical University in Prague, B\v{r}ehov\'{a} 7, 11519 Prague 1, Czech Republic \label{addr3}
           \and
           Institute of Particle Physics and Key Laboratory of Quark and Lepton Physics (MOE), Central China Normal University, Wuhan, 430079, China \label{addr4}
           \and
           Helmholtz Research Academy Hesse for FAIR (HFHF), GSI Helmholtz Center,
           Campus Frankfurt, Max-von-Laue-Stra{\ss}e 12, 60438 Frankfurt am Main, Germany \label{addr6}
}

\date{Received: date / Accepted: date}

\maketitle

\sloppy
\begin{abstract}
  Heavy-ion collisions at varying collision energies provide access to different regions of the QCD phase diagram. In particular collisions at intermediate energies are promising candidates to experimentally identify the postulated first order phase transition and critical end point.
  While heavy-ion collisions at low and high collision energies are theoretically well described by transport approaches and hydrodynamics+transport hybrid approaches, respectively, intermediate energy collisions remain a challenge.
  In this work, a modular hybrid approach, the \texttt{SMASH-vHLLE-hybrid} coupling 3+1D viscous hydrodynamics (\texttt{vHLLE}) to hadronic transport (\texttt{SMASH}), is introduced. It is validated and subsequently applied in Au+Au/Pb+Pb collisions between \sqrts = 4.3 GeV and \sqrts = 200.0 GeV to study the rapidity and transverse mass distributions of identified particles as well as excitation functions for \midyyield and $\langle p_\mathrm{T} \rangle$. A good agreement with experimental measurements is obtained, including the baryon stopping dynamics. The transition from a Gaussian rapidity spectrum of protons at lower energies to the double-hump structure at high energies is reproduced. The centrality and energy dependence of charged particle $v_2$ is also described reasonably well. This work serves as a basis for further studies, e.g. systematic investigations of different equations of state or transport coefficients.
\keywords{heavy-ion collisions \and hybrid approach \and hadron gas equation of state}
\end{abstract}

\section{Introduction}
\label{sec:intro}
Collisions of atomic nuclei at relativistic velocities provide a unique opportunity to study and understand the fundamental properties of matter \cite{Busza:2018rrf}. Such heavy-ion collisions are realized at different facilities, including for example the Relativistic Heavy-Ion Collider (RHIC) at BNL, the Large Hadron Collider (LHC) at Cern and SIS-18 at GSI.
These different facilities provide, by virtue of distinct experiment geometries and accelerator dimensions (collision energies), access to different regions of the QCD phase diagram, ranging from high temperatures and vanishing baryon chemical potentials to moderate temperatures and high baryon chemical potentials.
In recent years, heavy-ion collisions at intermediate energies have received particular attention in the quest for signatures of the postulated first order phase transition and critical end point \cite{Stephanov:1998dy, Asakawa:1989bq}. In particular the BESII program at BNL \cite{Kumar:2013cqa}, the NA61/SHINE experiment at Cern \cite{Turko:2018kvt} as well as future FAIR \cite{Wilczek:2010ae}, NICA \cite{Kekelidze:2016wkp} and J-PARC-HI \cite{Sako:2014fha} facilities have already or will soon provide experimental measurements in this designated region of interest \cite{Friese:2019qyw}. \\
There is a large variety of models to theoretically describe the underlying dynamics of heavy-ion collisions.
At low collision energies, where the system is dominated by hadronic interactions, the dynamics are successfully captured by transport approaches relying on hadronic degrees of freedom \cite{Weil:2016zrk, Buss:2011mx, Ehehalt:1996uq, Nara:1999dz, Bass:1998ca}. At high collision energies, where a quark-gluon plasma state of matter becomes accessible, so-called hybrid models constitute the standard approach \cite{Shen:2021nbe, Shen:2014vra, Bernhard:2016tnd, Schenke:2020mbo, JETSCAPE:2020mzn}. Relativistic (viscous) hydrodynamic calculations are coupled to a hadronic afterburner; a setup successfully reproducing experimental observables \cite{Petersen:2014yqa}. For heavy-ion collisions at intermediate energies, where baryon densities are finite, there is however no such standard description yet. Previous works relying on different initial conditions, hydrodynamic evolutions and hadronic transport approaches include for example \cite{Karpenko:2015xea, Petersen:2008dd, Wu:2021fjf, Shen:2017ruz, Bass:2000ib, Hirano:2012kj, Nonaka:2005aj, Shen:2014vra}.
These have however either not been applied below \sqrts = 7.7 GeV, rely on relativistic hydrodynamics in 2+1D, or have issues properly reproducing baryon dynamics at low collision energies. The proper description of the baryon stopping dynamics is crucial for studies concerning fluctuations of conserved charges and to properly model the evolution at intermediate beam energies.
Recent progress has been made in \cite{Akamatsu:2018olk, Du:2018mpf, Shen:2017bsr}, where a framework towards more dynamical initialization of the hydrodynamic phase has been introduced. This is of importance for lower collision energies due to the longer-lived initial state. \\

In this work, the \hybrid is introduced. It is conceptually similar to the hybrid models presented in \cite{Karpenko:2015xea, Petersen:2008dd} but comprises of an improved description of the baryon stopping dynamics towards lower collision energies.
The \hybrid relies on \smsh (Simulating Many Accelerated Strongly-Interacting Hadrons) \cite{Weil:2016zrk, SMASH_github} for the initial and the hadronic rescattering stage and on \vhlle \cite{Karpenko:2013wva, vhlle_github} for the hydrodynamic evolution.
It can be applied to model heavy-ion collisions ranging from \sqrts = 4.3 GeV to \sqrts = 5.02 TeV. It was already successfully employed to study the annihilation and regeneration of (anti-)protons and (anti-)baryons at a variety of collision energies and centralities \cite{Garcia-Montero:2021haa}. Here, the details of the \hybrid are explained, the approach is validated and results for particle spectra and excitation functions confronted with experimental data, where a good agreement across a wide range of collision energies is observed. In particular, the proton \dndy spectra as well as the proton \meanpt excitation function agree well with experimental measurements. \\

This work is structured as follows: In Sec.~\ref{sec:model_description} the \hybrid is explained with particular emphasis on the individual modules. The equation of state of the \smsh hadron resonance gas, which is of fundamental importance for the particlization process, is presented in Sec.~\ref{sec:SMASH-EoS}.
The approach is validated in Sec.~\ref{sec:validation} by demonstrating consistency at the interfaces, global conservation of quantum numbers and good agreement with other hybrid approaches consisting of \texttt{SMASH}+\texttt{CLVisc} \cite{Wu:2021fjf} and \texttt{vHLLE}+\texttt{UrQMD} \cite{Karpenko:2015xea}.
In Sec.~\ref{sec:Results}, results obtained with the \hybrid are presented. First, \dndy and \dndmt spectra for $\pi^-, p$ and $K^-$ in Au+Au/Pb+Pb collisions ranging from \sqrts = 4.3 GeV to \sqrts = 200.0 GeV as well as excitation functions for \midyyield and \meanpt are confronted with experimental data. A good agreement is obtained for pions, kaons and also the usually challenging protons. Second, the centrality and collision energy dependence of the charged particle integrated $v_2$ is well reproduced.
To conclude, a brief summary and outlook are provided in Sec.~\ref{sec:conclusions}.

\section{Model Description}
\label{sec:model_description}
The \hybrid \cite{hybrid_github} presented in this work is a novel hybrid approach for the theoretical description of heavy-ion collisions  between \sqrts = 4.3 GeV and \sqrts = 5.02 TeV. It is publicly available on Github\footnote{https://github.com/smash-transport/smash-vhlle-hybrid}. In general, hybrid approaches are a very successful combination of viscous fluid dynamics for the hot and dense region, while microscopic non-equilibrium transport approaches are applied for the initial and final stages where the system is far away from local equilibrium. In this work, we rely on the same transport approach (\texttt{SMASH}) for initial and final stages to allow for a smooth transition from the low energy regime, where hadronic transport approaches describe the full evolution to the higher energy region, which is where the phase transition to the quark-gluon plasma is expected.
The \hybrid consists of multiple submodules: The initial state is provided by the hadronic transport approach \smsh \cite{Weil:2016zrk, SMASH_github}, while the evolution of the hot and dense fireball is described by \texttt{vHLLE}, a 3+1D viscous hydrodynamics model \cite{Karpenko:2013wva, vhlle_github}. The hydroynamical evolution is performed until the medium gets too dilute, estimated with the energy density $e$ falling below a critical value of $e_{\mathrm crit} = 0.5$ GeV/fm$^3$. Particlization is realized with the \sampler\cite{sampler_github} and the subsequent hadronic afterburner evolution is again performed with the \smsh transport approach.
These submodules are explained in more detail in the following.

\subsection{\smsh}
\label{sec:SMASH}
\smsh is a hadronic transport approach designed for the description of low-energy heavy-ion collisions and hadronic matter in and out of equilibrium \cite{Weil:2016zrk, SMASH_github}. It provides an effective solution of the relativistic Boltzmann equation
\begin{align}
  p_\mu \ \partial^\mu \ f + m \ \partial_{p_\mu} \ (F^\mu \ f) \ = \ C(f)
  \label{eq:Boltzmann}
\end{align}
by modeling the collision integral $C(f)$ through formations, decays and elastic scatterings of hadronic resonances. In Eq.~(\ref{eq:Boltzmann}), $f$ denotes the one-particle distribution function, $p_\mu$ the particle's 4-momentum, $m$ its mass and $F^\mu$ an external force. The \smsh degrees of freedom consist of all hadrons listed by the PDG up to a mass of $m \approx$ 2.35 GeV \cite{Tanabashi:2018oca}. Resonances are represented by vacuum Breit-Wigner spectral functions, but have mass-dependent widths following the Manley-Saleski ansatz \cite{Manley:1992yb}. The resonance properties are tuned to reproduce elementary cross sections.
For high-energy hadronic interactions, a string model is employed where string fragmentation and the high energy hard scatterings are carried out within \texttt{Pythia 8} \cite{Sjostrand:2006za, Sjostrand:2007gs}.
In addition, \smsh has the perturbative production of dileptons \cite{Staudenmaier:2017vtq} and photons \cite{Schafer:2019edr} integrated. In this work, a geometric collision criterion is applied that is formulated in a covariant form \cite{Hirano:2012yy}, a stochastic collision criterion is however also implemented \cite{Staudenmaier:2020xqr}.
Inter alia, \smsh was already successfully applied to study particle production in a variety of collision setups \cite{Weil:2016zrk, Mohs:2019iee, Steinberg:2019wgm, Staudenmaier:2020xqr}. For a broader introduction into \texttt{SMASH} and its different features, the interested reader is referred to \cite{Weil:2016zrk, Mohs:2019iee}. In this work, the nuclear mean fields are turned off and only the cascade mode of \texttt{SMASH} is applied, which is appropriate for the beam energies under consideration.

\subsubsection{Initial conditions}
\label{sec:IC_SMASH}
The initial conditions for the hydrodynamical evolution are extracted from \smsh on a hypersurface of constant proper time $\tau_0$. By default, this proper time corresponds to the passing time of the two nuclei \cite{Karpenko:2015xea} and is determined from nuclear overlap via
\begin{align}
    \tau_0 =\frac{R_p + R_t}{\sqrt{\left(\frac{\sqrt{s_\mathrm{NN}}}{2 \ m_\mathrm{N}}\right)^2 - 1}},
  \label{Proper_time_IC}
\end{align}
where $R_p$ and $R_t$ are the radii of projectile and target, respectively.
$\sqrt{s_\mathrm{NN}}$ is the collision energy and $m_\mathrm{N}$ is the nucleon mass. The passing time constitutes a reasonable assumption for the time where local equilibrium is reached \cite{Oliinychenko:2015lva} and for high beam energies a lower limit of $\tau_0=0.5$ fm/c is implemented. To determine the iso-$\tau$ hypersurface, the \smsh evolution, which is realized in Cartesian coordinates, is initialized and performed including all scatterings and interactions until the particles reach the iso-$\tau$ hypersurface. Upon crossing this hypersurface, they are removed from the evolution and stored such that they can be later utilized to initialize the subsequent hydrodynamical evolution.
One might argue that the removal of particles introduces holes in the density distribution of the medium and could therefore modify the underlying dynamics. We have however verified that the final state observables remain unaffected if the particles are not removed from the \smsh evolution but can undergo further interactions.

\subsubsection{Hadronic Rescattering}
\label{sec:Afterburner_SMASH}
In the last stage of the \texttt{SMASH-vHLLE-hybrid}, \smsh is again applied to model the non-equilibrium hadronic afterburner evolution. For this, the hadrons obtained from particlization are read from an external file, back-propagated to the earliest time and successively appear once fully formed. They free stream before reaching their respective formation times. The afterburner stage is characterized by hadronic rescatterings as well as resonance decays. Both kinds of interaction take place until the medium is too dilute.

\subsection{\vhlle}
\label{sec:vhlle}
The evolution of the hot and dense fireball is modeled by means of the 3+1D viscous hydrodynamics code \vhlle \cite{Karpenko:2013wva}. It provides a solution of the hydrodynamic equations
\begin{align}
  \partial_\nu T^{\mu\nu} &= 0
  \label{eq:dmuTmunu}
  \\
  \partial_\nu N_\mathrm{B}^{\ \nu} = 0 \qquad
  \partial_\nu N_\mathrm{Q}^{\ \nu} &= 0 \qquad
  \partial_\nu N_\mathrm{S}^{\ \nu} = 0
  \label{eq:dmuNmu}
\end{align}
with the energy-momentum tensor decomposed as follows:
\begin{align}
T^{\mu\nu}=\epsilon u^\mu u^\nu - \Delta^{\mu\nu}(p+\Pi) + \pi^{\mu\nu} ,
\end{align}
where $\epsilon$ is the local rest frame energy density, $p$ and $\Pi$ are the equilibrium and bulk pressure, and $\pi^{\mu\nu}$ is the shear stress tensor.
Eqs.~(\ref{eq:dmuTmunu}) and (\ref{eq:dmuNmu}) encapsulate the conservation of energy and momentum via the energy momentum tensor $T^{\mu\nu}$ as well as the conservation of net-baryon, net-charge and net-strangeness numbers $N_\mathrm{B}$, $N_\mathrm{Q}$, $N_\mathrm{S}$, respectively.
The code solves relativistic viscous hydrodynamic equations in the second-order Israel-Stewart framework in the 14-momentum approximation, including shear-bulk coupling terms \cite{Denicol:2014vaa, Ryu:2015vwa}. That yields the following relaxation type equations for the viscous corrections:
\begin{align}
    D\Pi=&\frac{-\zeta\theta-\Pi}{\tau_\Pi}-\frac{\delta_{\Pi\Pi}}{\tau_\Pi}\Pi\theta+\frac{\lambda_{\Pi\pi}}{\tau_\Pi}\pi^{\mu\nu}\sigma_{\mu\nu},\label{equation:IS1}
    \\
    D\pi^{\langle\mu\nu\rangle} =& \frac{2\eta\sigma^{\mu\nu}-\pi^{\mu\nu}}{\tau_\pi}-
   \frac{\delta_{\pi\pi}}{\tau_\pi}\pi^{\mu\nu}\theta + \frac{\phi_7}{\tau_\pi}\pi_\alpha^{\langle\mu}\pi^{\nu\rangle\alpha} \nonumber
   \\
   &-\frac{\tau_{\pi\pi}}{\tau_\pi}\pi_\alpha^{\langle\mu}\sigma^{\nu\rangle\alpha} + \frac{\lambda_{\pi\Pi}}{\tau_\pi}\Pi\sigma^{\mu\nu}.\label{equation:IS2},
\end{align}
where $\theta$ is expansion scalar, $\sigma^{\mu\nu}$ is the stress tensor, $\tau_\pi$ and $\tau_\Pi$ are the relaxation times for the shear and bulk corrections and $\lambda_{\Pi\pi}$, $\lambda_{\pi\Pi}$, $\delta_{\pi\pi}$, $\tau_{\pi\pi}$ and $\phi_7$ are higher-order couplings, whose ansatzes are taken from \cite{Denicol:2014vaa}.
In this work, we only apply fixed values of effective shear viscosity over entropy density for the hydrodynamic evolution.
\vhlle is a Godunov-type algorithm in conservative form. In its original form, in Cartesian coordinates, the algorithm preserves the total energy, momentum and conserved charges in the system by construction. However, hydrodynamic equations in Milne coordinates contain non-vanishing source terms, which are integrated numerically with finite accuracy. Therefore, in Milne coordinates the total energy and momentum of the system varies a little, typically increasing from the beginning of the hydrodynamic evolution towards its end by few percents \cite{Schafer:2021rfk}. \\

The hydrodynamical evolution is initialized with the energy-momentum tensor calculated from the iso-$\tau$ particle list created with \smsh as detailed in Sec.~\ref{sec:IC_SMASH}. To prevent shock waves, these particles are smeared relying on a Gaussian smearing kernel as described in detail in \cite{Karpenko:2015xea}. The magnitude of the smearing is controlled by $R_\eta$ and $R_\perp$, the longitudinal and transversal smearing parameters, respectively. The specific values used in this work are provided in Sec.~\ref{sec:hybrid_configuration}. Once the particles are transformed into fluid elements, the medium is evolved relying on a chiral model equation of state \cite{Steinheimer:2010ib} which is matched to a hadron resonance gas at low energy densities.
The equation of state \cite{Steinheimer:2010ib} is constructed for isospin symmetric matter, i.e.\ zero electric charge chemical potential $\mu_Q=0$. Nevertheless, the electric charge density is included in the simulation. As such, the density is computed in the initial state and is evolved in the hydrodynamic stage. At particlization, a hadron gas equation of state is used, which includes both charges and corresponding chemical potentials and as such is not limited to isospin symmetric matter. In contrast to the baryon and electric charge densities, the strangeness density is set to zero in the hydrodynamic stage, as explained in Sec.~\ref{sec:SMASH-EoS}. \\
The viscous evolution is performed until the medium drops below the predefined critical energy density of $e_\mathrm{crit} = 0.5$ GeV/fm$^3$ and the corresponding freezeout hypersurface is created.
The latter is constructed dynamically during the hydrodynamical evolution with the \texttt{CORNELIUS} subroutine \cite{Huovinen:2012is}. For the sake of quantum number conservation in the subsequent particlization process, it is important that the thermodynamical properties of the hypersurface elements match those of the \smsh hadron resonance gas \cite{Schafer:2021rfk}. The \smsh equation of state is thus used to re-calculate the temperature and chemical potentials at each element. It is introduced in Sec.~\ref{sec:SMASH-EoS}.
The hydrodynamical evolution terminates once all cells have dropped below the critical energy density $e_\mathrm{crit} = 0.5$ GeV/fm$^3$. An energy density criterion results in a reasonable transition region in the QCD phase diagram.
For a more detailed introduction into \vhlle as well as its initialization from an external particle list, the reader is referred to \cite{Karpenko:2013wva, Karpenko:2015xea}.

\subsection{\sampler}
 \label{sec:sampler}
To perform the non-equilium hadronic afterburner evolution it is necessary to transform the fluid elements on the freezeout hypersurface into particles, a process usually referred to as particlization. In the presented model, the \texttt{SMASH-hadron-sampler} \cite{sampler_github} is applied to achieve particlization
according to the properties of the \smsh hadron resonance gas. Grand-canonical ensemble is employed, which allows to particlize each surface element independently. Technically this is realized in two steps: first, the sum of thermal multiplicities $N^{\rm tot}_i$ of all hadron species\footnote{Note though, that we exclude the leptons $e^\pm, \mu^\pm, \tau^\pm$, the photon $\gamma$ and the $\sigma$ meson from the \smsh particle list for the sampling process. These are not among the usual hadronic degrees of freedom of \smsh, but form part of the particle list to allow for additional features. They are also excluded for the determination of the \smsh equation of state in Sec.~\ref{sec:SMASH-EoS}.} passing through each surface element $i$ is computed. Next, an actual number of hadrons to generate at each surface element $N^{\rm gen}_i$ is sampled according to Poisson distribution with a mean $N^{\rm tot}_i$. If, for a given surface element $N^{\rm gen}_i>0$ then the generation proceeds to the second step: the type of a hadron is sampled based on the relative abundances from the thermal average, and the Cooper-Frye formula \cite{Cooper:1974mv} is applied to sample the momentum of each of the $N^{\rm gen}_i$ hadrons:
\begin{align}
  \begin{split}
    \label{eq:cooper-frye}
     \frac{\mathrm{d}N}{\mathrm{d}\vec{p}} = \frac{d}{(2\pi)^3} \  \int_\Sigma \left[ f_0(x, \vec{p}) + \delta f_\mathsf{shear}(x, \vec{p}) \ + \right. \\
     \left. \delta f_\mathsf{bulk}(x, \vec{p}) \right] \ \frac{p^\mu \ \mathrm{d} \Sigma_\mu}{E_{\vec{p}}} \qquad
  \end{split}
\end{align}
In Eq.~(\ref{eq:cooper-frye}), $N$ denotes the species' abundance, $d$ its degeneracy factor, $\vec{p}$ its 3-momentum, $p^\mu$ its 4-momentum, and $E_{\vec{p}}$ its energy. Furthermore, $\mathrm{d} \Sigma_\mu$ is the normal vector to the hypersurface patch, $f_0$ is the vacuum one-particle distribution function and $\delta f_\mathsf{shear}$ and $\delta f_\mathsf{bulk}$ are the viscous and bulk corrections to $f_0$, respectively. For the presented study, only shear viscous corrections are considered though.
The integration is performed over the full freezeout surface $\Sigma$ to account for all contributions from the individual hypersurface patches.
For further information about the specific sampling procedure, the reader is referred to Section 2 C of \cite{Karpenko:2015xea}.
Note that, in the \texttt{SMASH-hadron-sampler}, because of the grand-canonical nature of the sampling procedure, quantum numbers are not conserved event-by-event, but only on average. Since we are only interested in averaged single-particle observables, this approximation is sufficient at this point.
The resulting particle list is subsequently used to initialize the non-equilibrium hadronic afterburner evolution as described in Section~\ref{sec:Afterburner_SMASH} to model the late and dilute hadronic rescattering stage.

\subsection{\smsh hadron resonance gas equation of state}
\label{sec:SMASH-EoS}
\begin{figure}
  \centering
 \includegraphics[width = 0.45\textwidth]{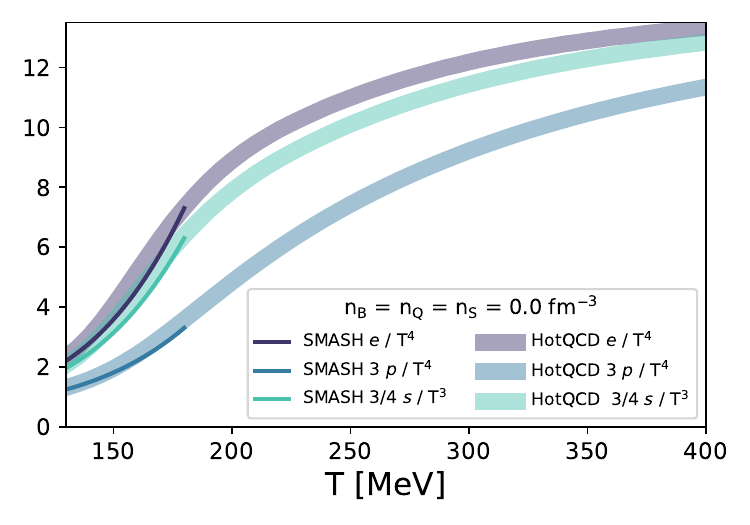}
 \caption{Equation of state of the \smsh hadron resonance gas in terms of the energy density $e$, pressure $p$, and entropy density $s$ (lines) in comparison to results obtained within 2+1-flavour lattice QCD  \cite{Bazavov:2014pvz} (bands) for vanishing net baryon, net charge and net strangeness densities.}
 \label{fig:EoS_Lattice}
\end{figure}

As described in Sec.~\ref{sec:vhlle}, the equation of state of the hadron gas corresponding to the degrees of freedom in the afterburner evolution is an essential ingredient for the creation of the freezeout hypersurface of the hydrodynamic stage. Otherwise, quantum number conservation in the particlization process is impossible. The equation of state of the hadron gas made up of the \smsh degrees of freedom is thus briefly described in the following. \\
The equation of state (EoS) generally provides a mapping of the thermodynamic quantities energy density $e$, net baryon density $n_\mathrm{B}$, net charge density $n_\mathrm{Q}$ and net strangeness density $n_\mathrm{S}$ to the temperature $T$, the pressure $p$, the baryon chemical potential $\mu_\mathrm{B}$, the charge chemical potential $\mu_\mathrm{Q}$, and the strange chemical potential $\mu_\mathrm{S}$. For heavy-ion collisions however, the net strangeness density can be approximated as \mbox{$n_\mathrm{S} = 0$ fm$^{-3}$}. We thus neglect the explicit dependence on the net strangeness density.
Whereas the hydrodynamic code is capable of evolving the strangeness current, in the present study $n_S=0$ is set in the initial state for the hydrodynamical stage. As such, local net strangeness is zero in the hydrodynamical as well as the particlization stage, which is consistent with the hadrnoic equation of state employed. Hence, the \smsh equation of state presented herein provides the mapping:
\begin{align}
  (e, n_\mathrm{B}, n_\mathrm{Q}) \ \to \ (T, p, \mu_\mathrm{B}, \mu_\mathrm{Q}, \mu_\mathrm{S})
\end{align}
This mapping follows from the properties of the underlying gas of hadrons and can be determined by solving the set of coupled equations
\begin{align}
  \begin{split}
    e &= e (T, \mu_\mathrm{B}, \mu_\mathrm{Q}, \mu_\mathrm{S}) \\
    n_\mathrm{B} &= n_\mathrm{B} \ (T, \mu_\mathrm{B}, \mu_\mathrm{Q}, \mu_\mathrm{S}) \\
    n_\mathrm{Q} &= n_\mathrm{Q} \ (T, \mu_\mathrm{B}, \mu_\mathrm{Q}, \mu_\mathrm{S}) \\
    n_\mathrm{S} &= n_\mathrm{S} \ (T, \mu_\mathrm{B}, \mu_\mathrm{Q}, \mu_\mathrm{S}),
  \end{split}
  \label{eq:EoS_equations}
\end{align}
assuming an ideal Boltzmann gas within the grand-canonical ensemble \cite{Oliinychenko:2016vkg}. The solutions of Eqs.~(\ref{eq:EoS_equations}) can numerically be obtained with a root solver algorithm. Unfortunately however, this solver is highly-sensitive to the choice of the initial approximation and might thus fail to converge.
This is particularly problematic at low energy densities, where the gas is loosely populated, and close to the kinematic thresholds\footnote{The kinematic thresholds for $n_\mathrm{B}$ and $n_\mathrm{Q}$ are defined from the composition of the gas, i.e. its lightest baryon and its lightest charged particle. In the case of the \smsh hadron resonance gas, the lightest baryon is the (anti-)proton with $m_\mathrm{p, \bar{p}} = 0.938$ GeV and the lightest charged particle is the pion with $m_\mathrm{\pi} = 0.138$ GeV. The kinematically accessible region is thus restricted to $e \geq m_\mathrm{p} \ |n_\mathrm{B}|$ and $e \geq m_\pi \ |n_\mathrm{Q}|$.}.
The \smsh equation of state presented herein is thus accurate for high energy densities as well as low baryon and charge densities, else it constitutes an approximation.
The latter is obtained from a combination of (i) variation of the grid size within the solver algorithm, (ii) variation of the initial approximations, and (iii) interpolations and extrapolations of the previously obtained solver results, where suitable.
The resulting equation of state, ranging from $e$ = 0.01 GeV/fm$^3$ up to $e$ = 1.0 GeV/fm$^3$, and accounting for all hadronic degrees of the \smsh hadron resonance gas, is available in  tabularized format under \cite{eos_github}. For validation purposes, the \smsh equation of state at vanishing net baryon, net electric charge and net strangeness densities is compared to lattice QCD results obtained in 2+1-flavour QCD \cite{Bazavov:2014pvz} in Fig.~\ref{fig:EoS_Lattice}.
\smsh results are denoted by lines, those from lattice QCD by bands. The energy density $e$, pressure $p$ and entropy density $s$, normalized to different orders of the temperature $T$, are in decent agreement with lattice QCD results at low temperatures, thus validating the hadron gas equation of state presented herein. \\
We want to mention that we have made different attempts to parametrize the \smsh equation of state, such that it could be applied more easily in this work, but also without much effort in other models. Unfortunately however, these attempts remain unsuccessful up to now. Nonetheless, a  brief summary of these endeavours is provided in \ref{app:parametrization}.

\subsection{Configuration details}
\label{sec:hybrid_configuration}
In this work, the \hybrid is applied for Au+Au and Pb+Pb collisions ranging from \sqrts = 4.3 GeV to \sqrts = 200.0 GeV. We simulate 100 event-by-event viscous hydrodynamics events, each initialized from one single \smsh event as described in Sec.~\ref{sec:IC_SMASH} and Sec.~\ref{sec:vhlle}. The shear viscosities and smearing parameters applied for the hydrodynamical evolution are listed in Table~\ref{tab:parameters}. For this first study, we chose to only employ a finite shear viscosity while setting the bulk viscosity to zero for the whole evolution.  Where applicable, the former are taken from \cite{Karpenko:2015xea}, else they are chosen to achieve a good description of the particle production.
From each resulting freezeout hypersurface 1000 events are sampled for the hadronic afterburner evolution. This allows for on-average quantum number conservation within the sampling process. \\
Unless stated differently, the code versions \linebreak \texttt{SMASH-vHLLE-hybrid:a1f823a}, \texttt{SMASH-2.0.2}, \linebreak \texttt{vHLLE:bce38e0}, \texttt{vhlle\_params:99ef7b4}, and \linebreak \texttt{SMASH-hadron-sampler-1.0} are used throughout this work.

\begin{table}
  \centering
  \begin{tabular}{c c c c c}
    System & \sqrts [GeV] & $\eta/s$ & $R_\perp$ & $R_\eta$ \\
    \hline
    Au + Au & 4.3 & 0.2 & 1.4 & 1.3 \\
    Pb+Pb & 6.4 & 0.2 & 1.4 & 1.2 \\
    Au + Au & 7.7 & 0.2 & 1.4 & 1.2 \\
    Pb+Pb & 8.8 & 0.2 & 1.4 & 1.0 \\
    Pb+Pb & 17.3 & 0.15 & 1.4 & 0.7 \\
    Au + Au & 27.0 & 0.12 & 1.0 & 0.4 \\
    Au + Au & 39.0 & 0.08 & 1.0 & 0.3 \\
    Au + Au & 62.4 & 0.08 & 1.0 & 0.6 \\
    Au + Au & 130.0 & 0.08 & 1.0 & 0.8 \\
    Au + Au & 200.0 & 0.08 & 1.0 & 1.0
  \end{tabular}
  \caption{Shear viscosities ($\eta / s$), transverse Gaussian smearing parameters ($R_\perp$), and longitudinal Gaussian smearing parameters ($R_\eta$) applied in this work for the hydrodynamical evolution of different collision systems and energies.}
  \label{tab:parameters}
\end{table}

\section{Model Validation}
\label{sec:validation}
Before applying the \hybrid to study different observables in heavy-ion collisions, we first want to systematically validate the approach. This is achieved with four different checks: First, the interfaces at the initial and final state of \vhlle are investigated with particular emphasis on the smeared initial state and the properties of the cells on the freezeout hypersurface, respectively. Second, the approximate global conservation of quantum numbers throughout the different stages of the evolution is demonstrated. Third, an apples-to-apples comparison to a hybrid model consisting of \smsh and \texttt{CLVisc} is performed, and fourth, results of the \hybrid are compared to results from a \texttt{UrQMD+vHLLE} hybrid model, as used in \cite{Karpenko:2015xea}. These validations and cross checks are elaborated on in more detail in the following.

\subsection{Interface with vHLLE}
\label{sec:interfaces}
It is well known that the interfaces between the non-equilibrium initial and final evolution and hydrodynamics are crucial in any hybrid approach. During the initial transition from \smsh to \vhlle the assumption of being close to local equilibrium leads to discontinuities. On the Cooper-Frye hypersurface, there is again a somewhat sharp transition between the strongly coupled fluid evolution and a hadronic transport approach with vacuum properties. Those fundamental problems are beyond the current work and are existent in all other hybrid approaches applied to heavy-ion collisions as well. Here, we are going to concentrate on ensuring minimal consistency in terms of matching quantum numbers at the initial interface. \\

\begin{figure}
  \centering
 \includegraphics[width = 0.45\textwidth]{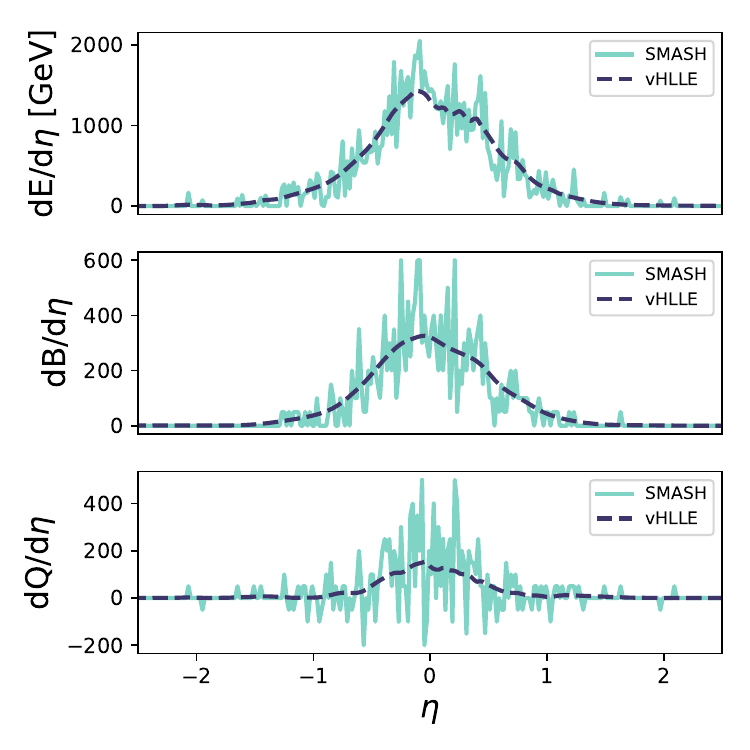}
\caption{Conserved quantities E, B and Q as a function of space-time rapidity $\eta$ at the initial \smsh $\to$ \vhlle interface for one single event of a Pb+Pb collision at \sqrts = 8.8 GeV. The \smsh state (solid lines) is compared to the smeared state used to initialize the hydrodynamical evolution in \vhlle (dashed line). Ideal hydrodynamics is applied and spectators are excluded. }
\label{fig:dEBQ_dEta}
\end{figure}

As described in Sec.~\ref{sec:vhlle}, it is necessary to smear the initial \smsh particle list upon initialization of \texttt{vHLLE}. The effect of this smearing is demonstrated in Fig.~\ref{fig:dEBQ_dEta}, containing the $\mathrm{d}E/\mathrm{d}\eta$ (upper), $\mathrm{d}B/\mathrm{d}\eta$ (center), $\mathrm{d}Q/\mathrm{d}\eta$ (lower) distribution as a function of space-time rapidity $\eta$.
The solid lines correspond to one single underlying \smsh event in a Pb+Pb collision at \sqrts = 8.8 GeV and the dashed line to its smeared counterpart, indicating the initial condition for the hydrodynamical evolution.
It can be observed that the spiky structure of the underlying \smsh event is smoothed successfully, without significant loss of quantum numbers, thus validating the handling of the \smsh $\to$ \vhlle interface. Fig. \ref{fig:dEBQ_dEta} gives also an idea on how large the event-by-event fluctuations in our initial state are.\\

\subsection{Global Conservation Laws}
\label{sec:conservation_laws}
The conservation of quantum numbers throughout all stages of the hybrid approach is essential in order to properly describe the evolution of the collision system. In this section we thus study the conserved quantities (energy $E$, net baryon number $B$ and net electric charge $Q$) as they evolve through the different modules and stages of the \texttt{SMASH-vHLLE-hybrid}.
\begin{figure*}
  \includegraphics[width = 0.32\textwidth]{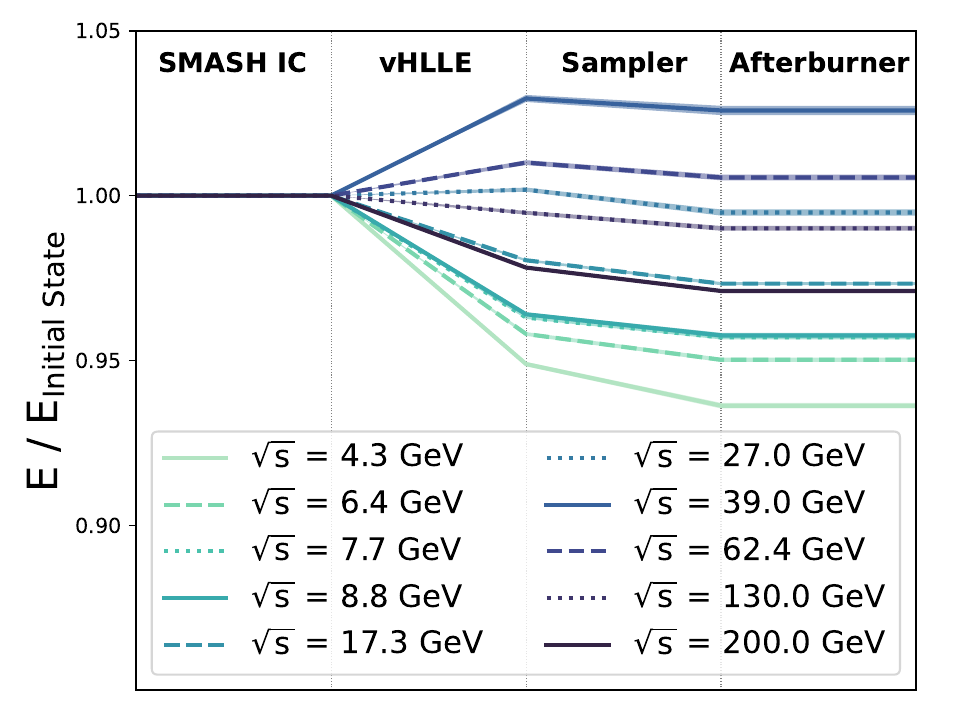}
  \hfill
  \includegraphics[width = 0.32\textwidth]{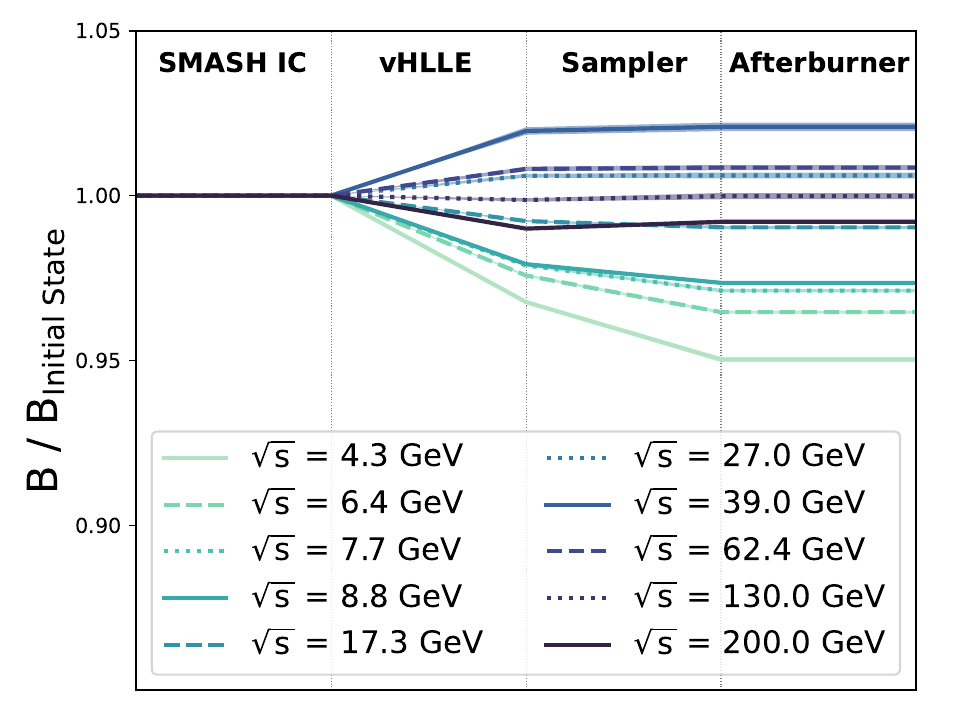}
  \hfill
  \includegraphics[width = 0.32\textwidth]{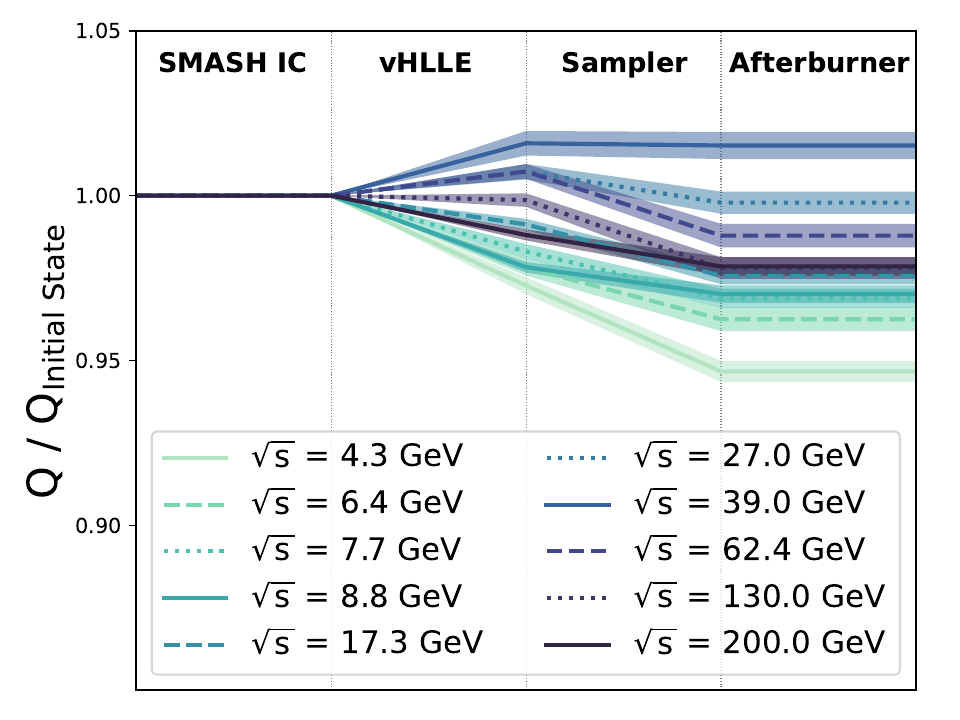}
  \caption{Conservation of total energy (left), baryon number (center), and electric charge (right) throughout all stages of the \hybrid for Au+Au and Pb+Pb collisions ranging from \sqrts = 4.3 GeV to \sqrts = 200.0 GeV. Note that for this conservation law study we rely on ideal instead of viscous hydrodynamics.}
  \label{fig:conserved_quantities}
\end{figure*}
Their evolution is presented in Fig.~\ref{fig:conserved_quantities}, where the left panel refers to the total energy $E$, the center panel to the net baryon number $B$, and the right panel to the net electric charge $Q$. These plots are to be interpreted as follows: The evolution of the conserved quantity from the initial to the final stage of the \hybrid is presented from left to right. The value displayed for $E, B$, or $Q$ is normalized to its respective initial value to identify deviations more easily. The segments of curves connect the normalized values at the beginning and the end of each stage in the modelling. As quantum number conservation is enforced in \smsh there are no violations up to the \smsh $\to$ \vhlle interface.
Deviations in the \vhlle stage represent a cumulative change to the quantum numbers from two sources. The first source is the finite accuracy of the hydrodynamic algorithm, due to the used Milne coordinate frame. In the Milne frame, geometrical source terms in the fluid-dynamical equations need to be time-integrated numerically, and the integration has a finite accuracy. The second source stems from the \smsh $\to$ \vhlle interface, however for visual purposes we combine it with the first source and display the result as an overall energy deviation during the \vhlle stage. At the initialization of the hydrodynmical evolution there is a small fraction of cells, typically located at the very borders of the fireball, whose energy density is already below $e \leq 0.01$ GeV/fm$^3$ and thus fall outside the applicability range of the \smsh equation of state (see Sect.~\ref{sec:SMASH-EoS}). Lacking possibilities to assign thermodynamic quantities ($T, p, \mu_\mathrm{B}, \mu_\mathrm{Q}, \mu_\mathrm{S}$) to these cells, we have decided to neglect their contribution, for the sake of not introducing additional uncertainties. Naturally, this treatment results in loss of $E$, $B$, and $Q$, which becomes more severe the lower the collision energy, as relatively more cells are characterized by small energy densities.
We have however verified, that at most 1.57 \% of the total energy, 0.74 \% of the total baryon number, and 0.29 \% of the total electric charge are lost in Au+Au collisions at \sqrts = 4.3 GeV.
Combining the afore-explained effects, the resulting violations in the \hybrid can be quantified to be of the order $\leq$ 5 \% up to completion of the \vhlle stage. In the subsequent sampling process, quantum numbers are approximately conserved. Only small violations are found, which are most pronounced at low collision energies. These violations are related to the fact that, as described in Sec.~\ref{sec:SMASH-EoS}, the hadron gas equation of state in this work is not perfectly accurate across the entire ($e, n_\mathrm{B}, n_\mathrm{Q}$)-range, but constitutes an approximation in the problematic regions. Finally, in the \smsh afterburner stage, quantum numbers are exactly conserved again, owing to its enforcement within the \smsh approach. Further discussions can be found in \cite{Schafer:2021rfk}.\\
Regarding the entire evolution of the system within the \texttt{SMASH-vHLLE-hybrid}, we find violations of quantum number conservation by at most 7 \% for collisions between \sqrts = 4.3 GeV and \sqrts = 200.0 GeV, which we consider an overall good validation.

\subsection{Comparison to CLVisc}
\label{sec:CLVisc_comp}
\begin{figure}
  \centering
 \includegraphics[width = 0.4\textwidth]{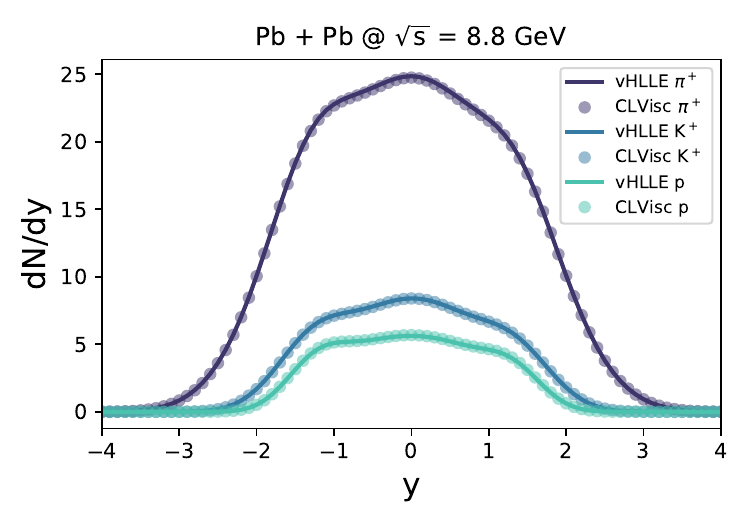}\\
  \includegraphics[width = 0.4\textwidth]{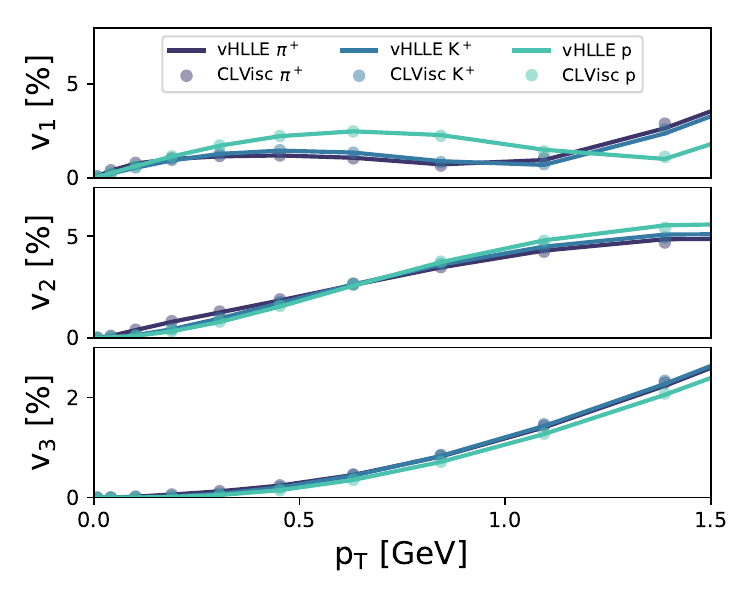}
\caption{Comparison of the dN/dy spectrum (upper) and anisotropic flow coefficients (lower) of pions, protons and kaons as obtained from directly evaluating the Cooper-Frye formula on the \vhlle and the \texttt{CLVisc} freezeout hypersurfaces. An identical \smsh initial state is used, as well as identical smearing parameters and equations of state.}
\label{fig:CLVisc_comp}
\end{figure}

The evolution up to the Cooper-Frye hypersurface is now validated by comparing its outcome to a hybrid model in which the hydrodynamical evolution is performed by \texttt{CLVisc} \cite{Pang:2018zzo, CLVisc_github} instead of \texttt{vHLLE}, both being 3+1D viscous hydrodynamics codes. For consistency, we use an identical initial state for a single event of a central Pb+Pb collision at a center-of-mass energy of $\sqrt{s}$ = 8.8 GeV. For an adequate comparison,  contributions in the corona region\footnote{The corona region denotes the collection of cells whose energy density is below the critical energy density already at the initialization of the hydrodynamical evolution. These cells are directly written to the freezeout hypersurface and are not further propagated in the hydrodynamical evolution.}
are neglected, since they are treated differently in \vhlle and \texttt{CLVisc}. For the Gaussian smearing, $R_\eta$ = 1.0 and $R_\perp$ = 1.4 are applied and for simplification ideal hydrodynamics with $\eta/s$ = 0 is used. Throughout the entire evolution a chiral model equation of state \cite{Steinheimer:2010ib} is used; we do not switch to the \smsh equation of state for the creation of the freezeout hypersurface. The latter is not necessary, as for this qualitative study we evaluate the Cooper-Frye formula directly on the freezout hypersurface and neither particlization, nor the afterburner evolution are performed. \\
In Figure~\ref{fig:CLVisc_comp} the \dndy spectra of pions, protons and kaons (upper) as well as their anisotropic flow coefficients $v_1$, $v_2$, and $v_3$ (lower) are compared between the hydrodynamical evolutions according to \vhlle and to \texttt{CLVisc}. Results obtained with \vhlle are denoted with solid lines, those obtained with CLVisc with filled circles.
We find a perfect agreement for the \dndy spectra as well as for the anisotropic flow coefficients between the \vhlle and \texttt{CLVisc} hydrodynamical evolution, relying on an identical initial state. This is considered another validation of the presented model, in particular regarding the hydrodynamical evolution. While both fluid dynamics codes have been very well tested, it is very nice to see that they lead to exactly the same results starting from the same \smsh initial condition.

\subsection{Comparison to UrQMD+vHLLE hybrid approach}
\label{sec:Karpenko_Comp}
To finally put the results obtained with the \hybrid into context with other state-of-the-art approaches, particle spectra are compared to those obtained within a hybrid model consisting of \texttt{UrQMD} and \texttt{vHLLE}, which was successfully applied to model heavy-ion collisions down to \sqrts = 7.7 GeV \cite{Karpenko:2015xea}.
\begin{figure}
  \centering
  \includegraphics[width = 0.41\textwidth]{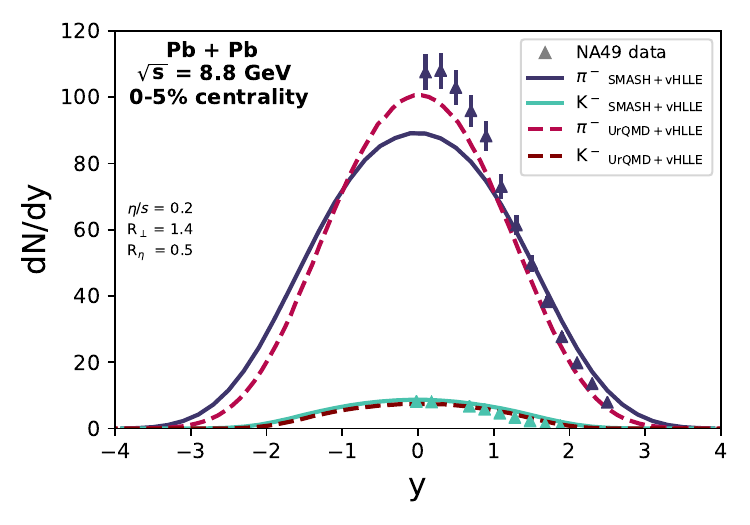}
  \hfill
  \includegraphics[width = 0.41\textwidth]{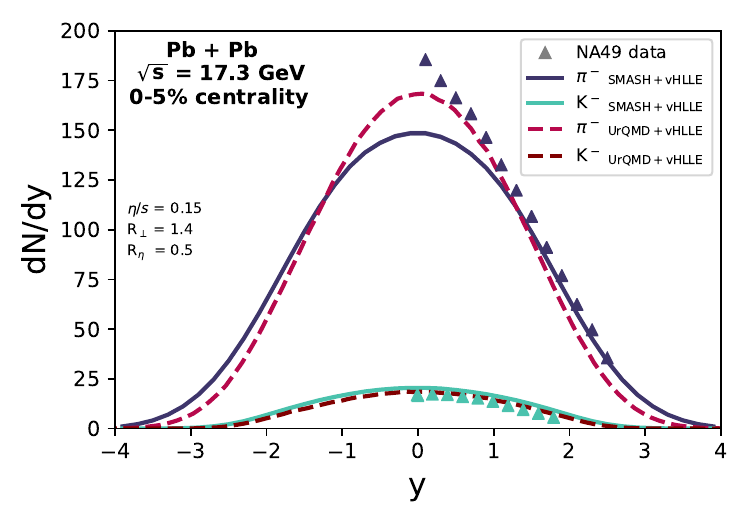} \\
  \includegraphics[width = 0.44\textwidth]{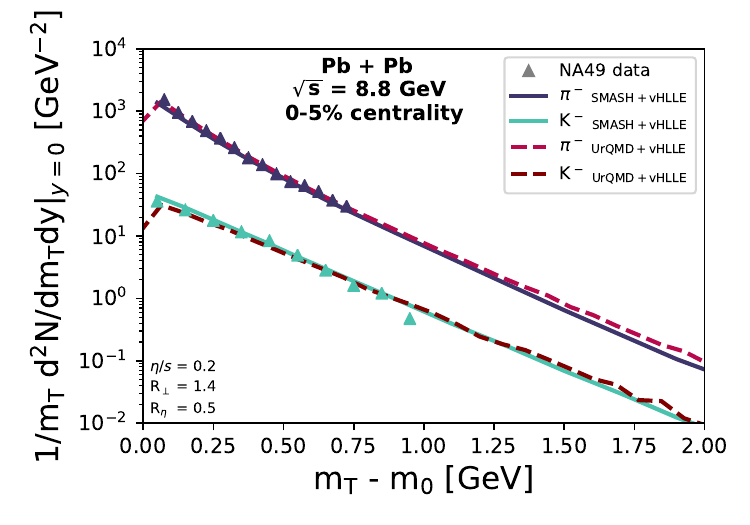}
  \hfill
  \includegraphics[width = 0.44\textwidth]{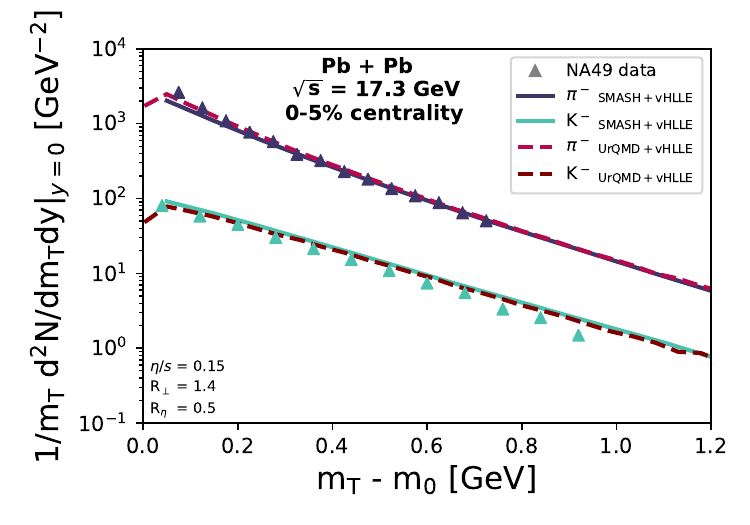}
  \caption{Comparison of the \dndy spectra (upper) and \mt spectra (lower) in central Pb+Pb collisions at \sqrts = 8.8 and 17.3 GeV between the \hybrid (solid lines) and the \texttt{UrQMD+vHLLE} hybrid (dashed) for $\pi^-$ and K$^-$. Results for the \texttt{UrQMD+vHLLE} hybrid are taken from \cite{Karpenko:2015xea}.
  Identical smearing parameters and viscosities are applied. The experimental NA49 data is taken from \cite{NA49:2002pzu, NA49:2005cix, NA49:2007stj, NA49:2010lhg}.}
  \label{fig:karpenko_comp}
\end{figure}

The conceptual ideas of the \hybrid and the \texttt{UrQMD+vHLLE} hybrid approach are identical. The only difference is the transport approach applied to provide the initial state and to perform the afterburner evolution. Although \texttt{UrQMD} and \texttt{SMASH} are conceptually similar, there are small differences regarding the degrees of freedom, cross sections, or in the application of \texttt{Pythia} to perform the string excitations. In particular the latter is important to properly model the baryon dynamics at intermediate and high collision energies \cite{Mohs:2019iee}. \\
Fig.~\ref{fig:karpenko_comp} shows the rapidity distributions \dndy (upper) and \mt spectra (lower) of $\pi^-$ and $K^-$ in Pb+Pb collisions at \sqrts = 8.8 GeV and 17.3 GeV as obtained within the \hybrid (solid lines) and within the \texttt{UrQMD+vHLLE} hybrid (dashed lines, taken from \cite{Karpenko:2015xea}). For consistency, identical smearing parameters and viscosities are applied to model the hydrodynamical evolution in both cases. Note that these differ from the ones listed in Table~\ref{tab:parameters}, the applied values are indicated directly in the figures. It can be observed that, at both collision energies, the \dndy distribution of pions from the \hybrid is wider than the one from the \texttt{UrQMD+vHLLE} hybrid, but peaks at lower midrapidity yields. The shape of the kaon \dndy distribution is similar, but the \texttt{UrQMD+vHLLE} hybrid produces smaller kaon yields than the \texttt{SMASH-vHLLE-hybrid}.
The \mt spectra of both approaches are in good agreement, only at low transverse masses the \texttt{UrQMD+vHLLE} hybrid produces more pions but less kaons than the \texttt{SMASH-vHLLE-hybrid}, which is not surprising given the differences in the \dndy spectra. It is reassuring to find that both hybrid approaches provide similar results in terms of \dndy and \mt spectra. This puts the novel hybrid approach introduced in this work into context with other approaches existent in the fied and gives some hints towards differences between the two transport approaches \smsh and \texttt{UrQMD}.

\section{Results}
\label{sec:Results}
In this section, bulk hadronic observables are presented as obtained from the \hybrid in heavy-ion collisions at different collision energies. This includes \dndy and \mt spectra of pions, protons and kaons as well as their excitation functions for the midrapidity yield, $\langle p_\mathrm{T} \rangle$, $v_2$, and $v_3$. The results from the hybrid approach are compared to a pure transport calculation to indicate the differences due to the intermediate hydrodynamic evolution. Before the final results for observables are discussed, we show results on the location of our switching transition between \vhlle and \smash on a hypersurface of constant energy density in the phase diagram. 

\subsection{Cooper-Frye transition hypersurface}

\begin{figure}
  \centering
 \includegraphics[width = 0.45\textwidth]{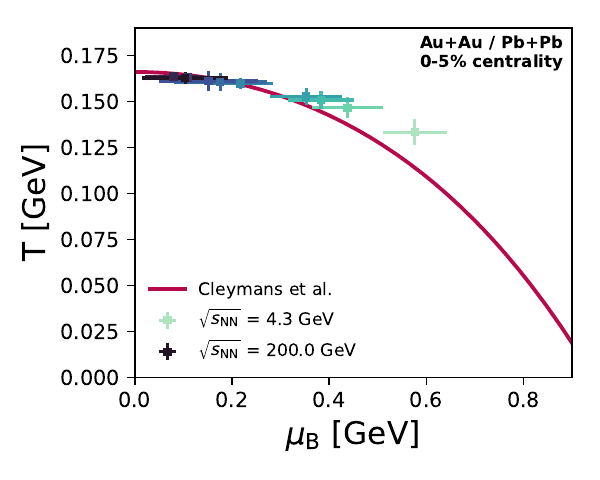}
\caption{Mean coordinates of the patches on the \vhlle Cooper-Frye hypersurface for central Au+Au/Pb+Pb collisions at \sqrts = 4.3 GeV (light blue), 6.4 GeV, 7.7 GeV, 8.8 GeV, 17.3 GeV, 27.0 GeV, 39.0 GeV, 62.4 GeV, 130.0 GeV, and 200.0 GeV (dark blue). The error bars correspond to the widths of the $T$ and $\mu_\mathrm{B}$ distributions, the solid line to a parametrization of experimentally obtained freezeout properties, taken from \cite{Cleymans:2006qe}.}
\label{fig:freezeout_diagram}
\end{figure}

The \vhlle $\to$ \sampler interface is examined by analyzing the properties of the Cooper-Frye hypersurface patches in the $T$-$\mu_\mathrm{B}$ plane. For this purpose, the expectation value for the properties of the switching hypersurface elements in terms of $T$ and $\mu_\mathrm{B}$ are determined in Au+Au/Pb+Pb collisions at a range of collision energies. Their locations in the $T$-$\mu_\mathrm{B}$ plane are depicted in Fig.~\ref{fig:freezeout_diagram}, where the differently coloured markers correspond to the results obtained within the \texttt{SMASH-vHLLE-hybrid} at varying collision energies. Lower collision energies are depicted with lighter colours, higher energies with darker colours.
The positions of the markers correspond to the mean, while the vertical and horizontal bars denote the width of the distribution in $T$ and in $\mu_\mathrm{B}$, respectively. The mean and standard deviation are obtained, weighted by the energy density of the respective hypersurface element:
\begin{align}
  \langle A \rangle &= \sum_{n=0}^{N} A_n\ \frac{e_n}{\sum_{n=0}^N e_n}  \\
  \sigma_A &= \sum_{n=0}^{N} \left( A_n - \langle A \rangle \right)^2 \ \frac{e_n}{\sum_{n=0}^N e_n}
\end{align}
Here, $A$ is a placeholder for the quantity of interest, $T$ or $\mu_\mathrm{B}$, $N$ is the total number of hypersurface elements, and $e_n$ is the energy density of the respective hypersurface element.
In Fig.~\ref{fig:freezeout_diagram}, the results for central collisions are presented, for a centrality-dependent version of this figure, the reader is referred to \ref{app:centrality_dep}.
The solid line in Fig.~\ref{fig:freezeout_diagram} corresponds to a parametrization of the chemical freezeout line, deduced from experimentally measured hadron abundances within the statistical hadronization model \cite{Cleymans:2006qe, Cleymans:1998fq}.
Regarding the locations of the Cooper-Frye transition surface within the \hybrid it can be observed that, as expected, higher-energy collisions are characterized by higher temperatures and lower baryon-chemical potentials, while lower collision energies by relatively lower temperatures and higher baryon-chemical potentials. This is generally in line with the shape of the parametrization.
Let us also point out here that the spread of the system is significant even in single events, so a large part of the phase diagram can be covered with a limited range of beam energies (see also \cite{Bass:2012gy}). As expected the transition from hydrodynamics to hadronic transport happens slightly above the chemical freeze-out line, since the system should be mainly hadronic at that stage. The actual chemical freezeout happens naturally within the rescattering dynamics, when the inelastic collisions cease depending on the hadronic species.

\subsection{Hadron production: \dndy and \dndmt}
\label{sec:spectra}
First, the \hybrid is applied to model Au+Au collisions at \sqrts = 4.3 GeV (E$_\mathrm{lab}$ = 8 AGeV) and at \sqrts = 7.7 GeV (E$_\mathrm{lab}$ = 30 AGeV), as well as Pb+Pb collisions at \sqrts = 17.3 GeV (E$_\mathrm{lab}$ = 158 AGeV). The viscosities and smearing parameters applied for the hydrodynamical evolution are those listed in Table~\ref{tab:parameters}. An impact parameter range of $b \in [0, 3.3]$ fm in the case of Au+Au collisions and of $b \in [0, 3.1]$ fm in the case of Pb+Pb collisions is used as a proxy for collisions in the 0-5\% centrality range.
\begin{figure*}
  \centering
  \includegraphics[width = 0.47\textwidth]{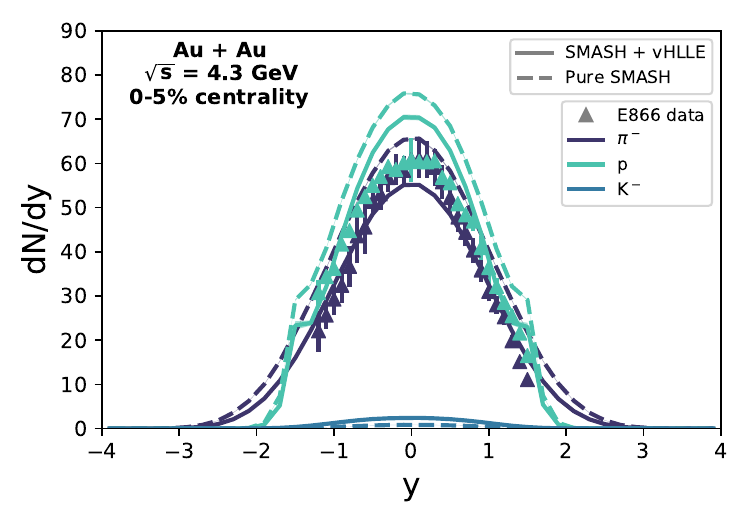}
  \hfill
  \includegraphics[width = 0.47\textwidth]{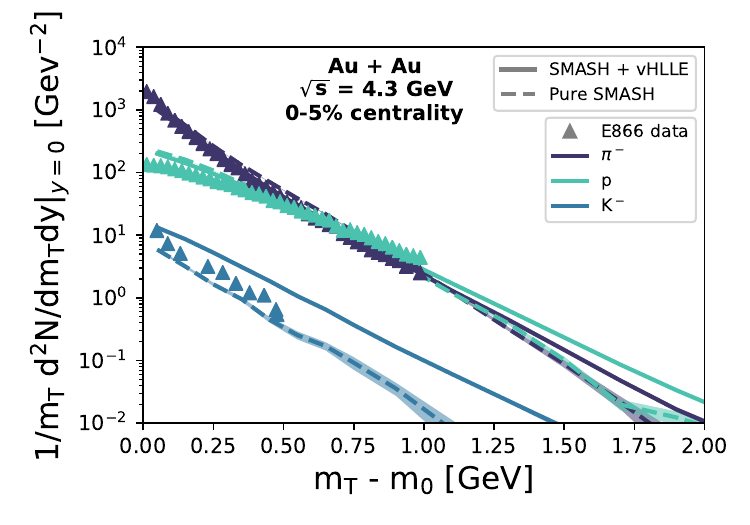}
  \includegraphics[width = 0.47\textwidth]{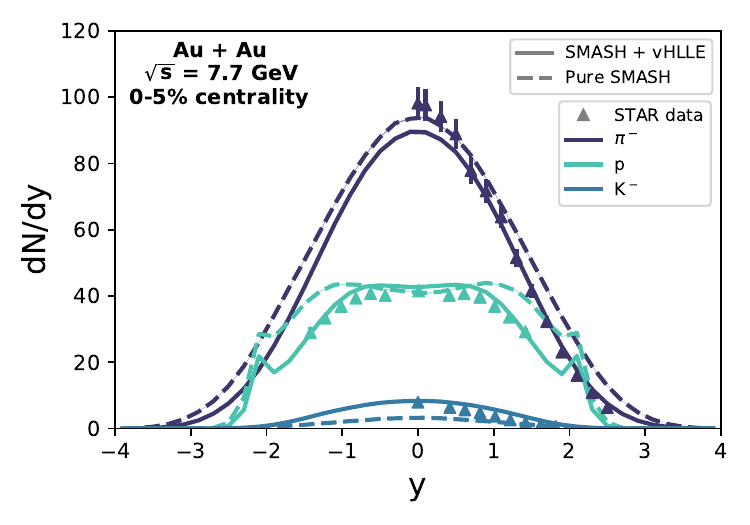}
  \hfill
  \includegraphics[width = 0.47\textwidth]{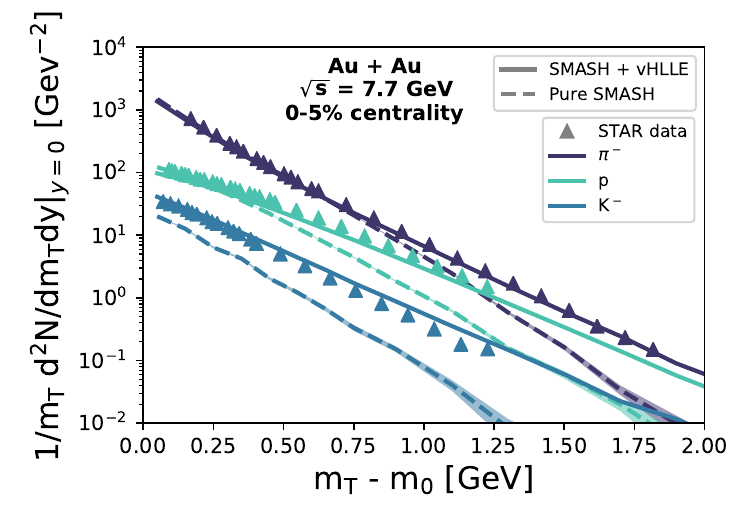}
  \includegraphics[width = 0.47\textwidth]{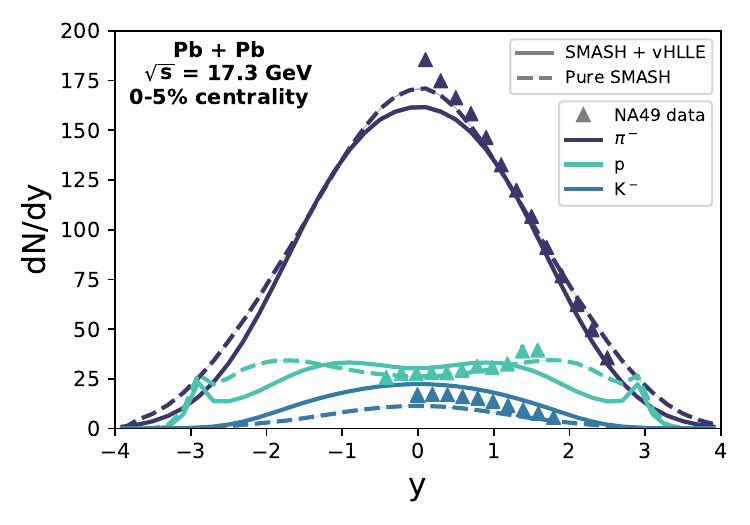}
  \hfill
  \includegraphics[width = 0.47\textwidth]{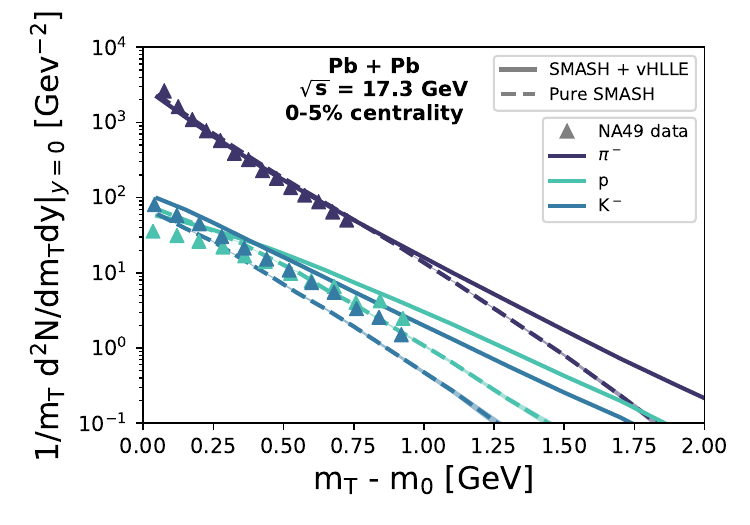}
  \caption{\dndy spectra (left) and \mt spectra (right)  of $\pi^-$, $p$, and $K^-$ for central Au+Au/Pb+Pb collisions at \sqrts = 4.3 GeV (upper), \sqrts = 7.7 GeV (center), and \sqrts = 17.3 GeV (lower). The results obtained within the \hybrid (solid lines) are compared to those obtained when running only \smsh (dashed lines).
  The E866 data is taken from \cite{E-0895:2003oas, Ahle:2000wq}, the STAR data from \cite{STAR:2017sal}, and the NA49 data from \cite{NA49:2002pzu, NA49:2005cix, NA49:2007stj, NA49:2010lhg}.}
  \label{fig:dndy_mT_spectra}
\end{figure*}
In Fig.~\ref{fig:dndy_mT_spectra} the \dndy spectra (left column) and \mt spectra (right column) of $\pi^-$, $p$, and $K^-$ in Au+Au/Pb+Pb collisions are presented for three different collision energies: \sqrts = 4.3 GeV (upper row), \sqrts = 7.7 GeV (center row), and \sqrts = 17.3 GeV (lower row). Results from the \hybrid are marked by solid lines and compared to those obtained from running \smsh only (dashed lines), without any intermediate hydrodynamical stage.
Where available, the results are confronted with experimental data from the E866 \cite{E-0895:2003oas, Ahle:2000wq}, STAR \cite{STAR:2017sal}, and NA49 \cite{NA49:2002pzu, NA49:2005cix, NA49:2007stj, NA49:2010lhg} collaborations. \\
Regarding the \dndy spectra it can be observed that, across all collision energies, the application of a hybrid model instead of a pure transport evolution decreases the pion yield and enhances the kaon production. 
This is in line with observations made in previous works \cite{Petersen:2008dd, Karpenko:2015xea, Akamatsu:2018olk}. It can further be observed that the longitudinal dynamics of protons are qualitatively captured properly across all three collision energies. While at low energies the single-peak structure in proton \dndy spectra persist, baryon transparency has a greater impact for rising collision energies \cite{Blume:2007kw}. The \hybrid is successful at describing the single-peak structure observed at \sqrts = 4.3 GeV as well as the double-peak structure at \sqrts = 7.7 GeV and \sqrts = 17.3 GeV. The new hybrid approach is thus able to reproduce the magnitude of baryon stopping at intermediate collision energies. The agreement with experimental data for proton \dndy spectra is improved significantly with the application of the \texttt{SMASH-vHLLE-hybrid} instead of a pure transport evolution, yet the final agreement is not perfect. Let us note in that respect that finite baryon diffusion will also affect the proton rapidity distribution, which is not yet included in the \hybrid \cite{Shen:2017bsr, Du:2018mpf}. Nevertheless, the qualitative (and almost quantitative) agreement with the proton rapidity distribution over a large range of beam energies is considered an important pre-requisite for further studies of phase transition signals.\\
In the case of the \dndmt spectra in Fig.~\ref{fig:dndy_mT_spectra}, a significant hardening of the $\pi^-$, $p$, and $K^-$ spectra is observed at all collision energies if the \hybrid is employed instead of a pure transport evolution utilizing \texttt{SMASH}. These results are again in line with observations made in other hybrid approaches\cite{Petersen:2008dd, Karpenko:2015xea}. The observed hardening of the spectra in the hybrid setup noticeably improve the agreement with experimental measurements as compared to a pure transport evolution at all collision energies probed in Fig.~\ref{fig:dndy_mT_spectra}, in particular regarding the slopes of the \mt spectra of all particle species. \\
Summarizing Fig.~\ref{fig:dndy_mT_spectra} it can be stated that the \hybrid approach is reasonably describing \dndy and \dndmt spectra of $\pi^-$, $p$, and $K^-$ in Au+Au/Pb+Pb collisions at \sqrts = 4.3, 7.7 and 17.3 GeV. A decent agreement with experimental measurements at these energies is obtained. Also, the baryon stopping power is well reproduced at the above listed energies, thus demonstrating that longitudinal baryon dynamics are captured properly.

\subsection{Excitation functions}
\label{sec:exc_funcs}
To investigate the particle production over a broader range of collision energies, the midrapidity yield and mean \pt are calculated for $\pi^-$, $p$ and $K^-$ and compared to the pure transport calculation. Reproducing the yield and the mean transverse momentum is almost equivalent to reproducing the full transverse mass spectra. Au+Au/Pb+Pb collisions are calculated between \sqrts = 4.3 GeV and \sqrts = 200.0 GeV, again relying on the viscosities and smearing parameters listed in Table~\ref{tab:parameters}. As above, central collisions are modeled where the impact parameter range for 0-5\% central collisions is approximated by $b \in [0, 3.3]$ fm and $b \in [0, 3.1]$ in Au+Au and Pb+Pb collisions, respectively.
For centrality-dependent excitation functions, the interested reader is referred to \ref{app:centrality_dep}. \\

\begin{figure}
  \centering
  \includegraphics[width = 0.47\textwidth]{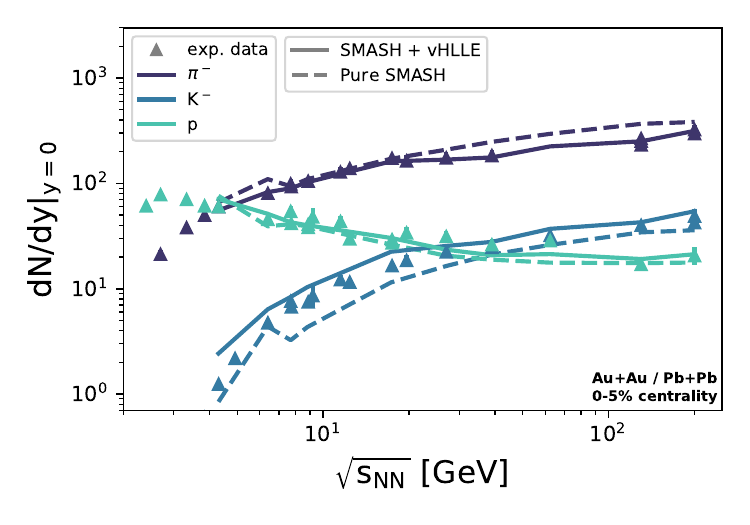}
  \includegraphics[width = 0.47\textwidth]{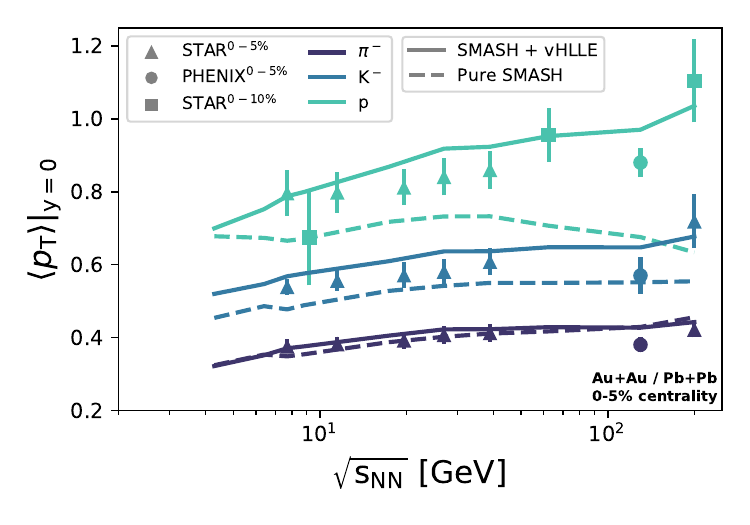}
  \caption{Midrapidity yield (upper) and mean \pt (lower) excitation functions as extracted from the \hybrid (solid lines) in comparison to a pure \smsh evolution (dashed lines) and to experimental data collected from
  \cite{E-0895:2003oas, NA49:2002pzu, STAR:2017sal, E895:2001zms, STAR:2009sxc, NA49:2006gaj, NA49:2010lhg, E866:2000dog, STAR:2009sxc}.}
  \label{fig:Midy_meanpT_exc_funcs}
\end{figure}
In Fig.~\ref{fig:Midy_meanpT_exc_funcs} the midraidity yield excitation function $\mathrm{d}N/\mathrm{d}y|_{y = 0}$ (upper) and the midrapidity mean-\pt excitation function $\langle p_\mathrm{T} \rangle |_{y = 0}$ (lower) are presented as a function of collision energy $\sqrt{s_\mathrm{NN}}$.
Results from the \hybrid are denoted with solid lines, those from applying a pure \smsh transport evolution with dashed lines.
The results are confronted with experimental data from the E895 \cite{E-0895:2003oas, E895:2001zms}, E866 \cite{E866:2000dog}, NA49 \cite{NA49:2002pzu, NA49:2006gaj, NA49:2010lhg}, and STAR \cite{STAR:2017sal, STAR:2009sxc, STAR:2009sxc} collaborations.
The $\mathrm{d}N/\mathrm{d}y|_{y = 0}$ excitation function determined within the \hybrid shows a good agreement with experimental measurements. There is a nearly-perfect agreement with the measured $\pi^-$ and $p$ yields, the $K^-$ yields are systematically overestimated though. The latter could be improved by more dynamical initial conditions, as recently realized in \cite{Akamatsu:2018olk}. It can further be observed that the excitation function from the \hybrid shows a smooth behaviour for rising collision energies, while a kink is observed between \sqrts = 6 GeV and \sqrts = 8 GeV in the pure \smsh curve. This kink is unphysical and stems from the non-trivial transition from resonance dynamics to string dynamics in \texttt{SMASH}. The \hybrid is thus better suited to smoothly and consistently describe the dynamics at intermediate collision energies.
Let us note here, that it is well known that the application of a pure transport model is not sufficient towards higher collision energies, as too little radial flow is produced. At the same time, pure hydrodynamics is not sufficient either for lack of non-equilibrium dynamics in the initial and final stages. In hydrodynamics+transport models, transport theory and hydrodynamics are employed in their respective regions of applicability to provide as good a description of heavy-ion collisions as possible. Ideally, unphysical properties arise only if these models are employed far from their region applicability.
 \\
For the $\langle p_\mathrm{T} \rangle |_{y = 0}$ excitation function displayed in the lower panel of Fig.~\ref{fig:Midy_meanpT_exc_funcs} the agreement with experimental measurements is also improved once a hybrid model is applied instead of a pure \smsh evolution. Most importantly, the \meanpt of protons rises with rising collision energies, which is in line with experimental observations, while in the pure \smsh case it shows a decreasing trend for collisions above \sqrts $\approx$ 30 GeV. The application of a hybrid model is thus essential to properly capture the transversal baryon dynamics, especially towards higher collision energies.

\subsection{Collective flow}
Third, the integrated elliptic flow $v_2$ and integrated triangular flow $v_3$ of charged particles are determined for
Au+Au/Pb+Pb collisions between \sqrts = 4.3 GeV and \sqrts = 200.0 GeV
as modeled with the \hybrid in 0-5\%, 5-10\%, 10-20\%, 20-30\%, 30-40\%, and 40-50\% most central collisions\footnote{The corresponding impact parameter ranges in Au+Au collisions are:
$b_{0-5\%} \in [0,3.3]$ fm, $b_{5-10\%} \in [3.3,4.6]$ fm, $b_{10-20\%} \in [4.6,6.5]$ fm, $b_{20-30\%} \in [6.5,7.9]$ fm, $b_{30-40\%} \in [7.9,8.5]$ fm, and $b_{40-50\%} \in [8.5,10.3]$ fm. \\
And those in Pb+Pb collisions:
$b_{0-5\%} \in [0,3.1]$ fm, $b_{5-10\%} \in [3.1,4.5]$ fm, $b_{10-20\%} \in [4.5,6.6]$ fm, $b_{20-30\%} \in [6.6,8.2]$ fm, $b_{30-40\%} \in [8.2,9.5]$ fm, and $b_{40-50\%} \in [9.5,10.6]$ fm.
}.
The event plane method is employed and the resulting $v_2$ and $v_3$ is compared to experimental data from the STAR collaboration \cite{STAR:2017idk}. Note though, that the STAR results are obtained from two-particle correlations, hence the comparison is not perfectly equivalent. It has been shown however that both methods yield similar, yet not identical, results \cite{STAR:2012och}.
\begin{figure}
  \centering
  \includegraphics[width = 0.47\textwidth]{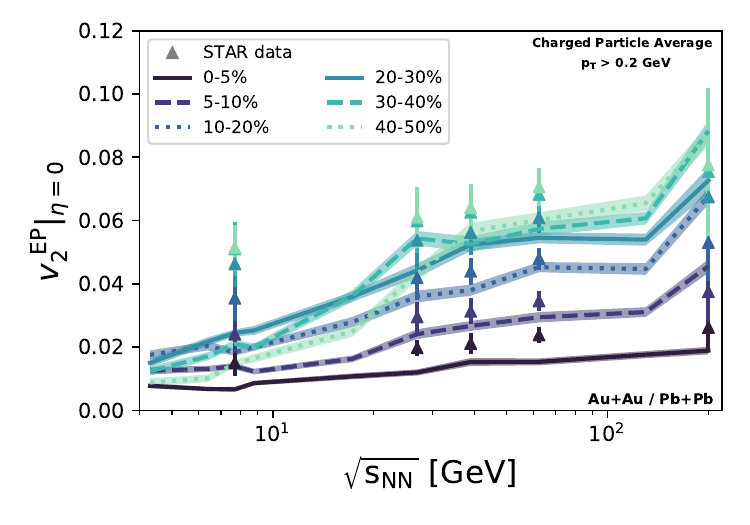}
  \includegraphics[width = 0.47\textwidth]{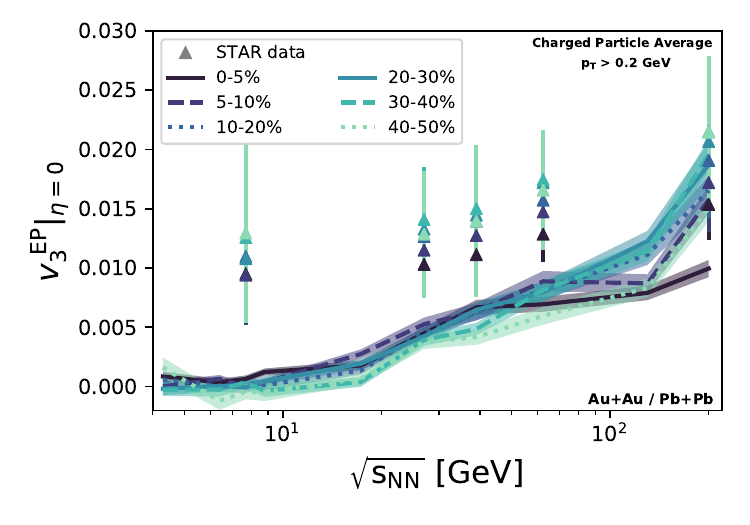}
  \caption{Integrated $v_2$ (upper) and integrated $v_3$ (lower) excitation functions of charged particles at midrapidity as a function of collision energy, obtained with the \hybrid in Au+Au/Pb+Pb collisions. More central collisions are marked by darker colours, more peripheral collisions by lighter colours.
  The STAR data is taken from \cite{STAR:2017idk}.}
  \label{fig:v2_v3}
\end{figure}
The integrated $v_2$ (upper panel) and integrated $v_3$ (lower panel) are presented as a function of collision energy \sqrts in Fig.~\ref{fig:v2_v3}.
More central collisions are marked by darker colours and more peripheral collision by lighter colours. \\
Regarding $v_2$, it can be observed that for central collisions the $v_2$ extracted from the \hybrid is in decent agreement with measurements from STAR, while for more peripheral collisions the agreement is good at high collision energies, but the measured $v_2$ is significantly underestimated towards lower collision energies. It can further be seen that the $v_2$ determined from the \hybrid in 20-30\%, 30-40\% and 40-50\% most central collisions is characterized by a clear kink around \sqrts $\approx$ 40.0 GeV and decreases much more than in more central collisions when moving from higher to lower collision energies.
This can be explained with a too short lifetime of the hot and dense fireball in the hydrodynamic stage.
\begin{figure}
  \centering
  \includegraphics[width = 0.47\textwidth]{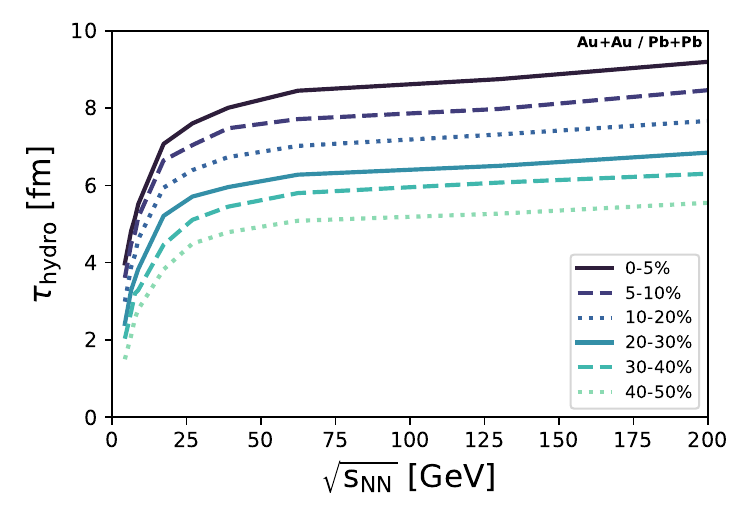}
  \caption{Lifetime of the hydrodynamically evolved fireball in Au+Au/Pb+Pb collisions within the \hybrid as a function of collision energy and for different centralities. More central collisions are marked by darker colours, more peripheral collisions by lighter colours.}
  \label{fig:lifetime_hydro}
\end{figure}
In Fig.~\ref{fig:lifetime_hydro} the mean lifetime of the hydrodynamic stage, $\tau_\mathrm{hydro}$, is presented as a function of collision energy for different centralities. Again, more central collisions are marked by darker colours and more peripheral collision by lighter colours. The hydrodynamical lifetime is determined via
\begin{align}
  \tau_\mathrm{hydro} = \langle \ \tau_\mathrm{end} \ \rangle_\mathrm{events} \ - \ \tau_0
\end{align}
where $\tau_0$ is the proper time at initialization of hydrodynamics and $\tau_\mathrm{end}$ the proper time at which the last cell falls below the critical energy density. $\langle \ \rangle_\mathrm{events}$ denotes the average over all events.
It can be seen that the lifetime of the fireball decreases continuously for increasing centralities. This is expected from the geometrically smaller overlap region resulting in smaller volumes and thus shorter-lived fireballs.
One can further observe that the hydrodynamical lifetime is nearly independent of the collision energy above \sqrts $\approx$ 50 GeV. Below \sqrts $\approx$ 50 GeV however, it is reduced substantially with decreasing collision energies.
This implies, that especially in peripheral collisions at low and intermediate collision energies, the hydrodynamical evolution lasts only a few fm/c, which is seemingly too short for collective dynamics to develop. The resulting $v_2$ is thus significantly underestimated.
This might be alleviated by implementing more dynamical initial conditions that are better suited to capture the underlying dynamics at low collision energies, as recently done in \cite{Akamatsu:2018olk}.\\
Regarding $v_3$ on the other hand (c.f. lower panel of Fig.~\ref{fig:v2_v3}), the experimentally measured $v_3$ is underestimated across the entire energy range and for all centralities by the \texttt{SMASH-vHLLE-hybrid}. This might be related to
the Gaussian smearing employed at initialization of hydrodynamics to provide smooth density profiles. Initial state fluctuations, which are responsible for the generation of $v_3$, might thus be smeared too much and the resulting $v_3$ is hence underestimated. Additionally, a too short hydrodynamical evolution, as detailed above, could also be responsible for an underestimation of the integrated $v_3$.

In general, collective flow observables are very sensitive to the presence of a fluid dynamic evolution and to the employed transport coefficients. It will be interesting to see, how these coefficients behave when more realistic temperature and net baryon density dependent transport coefficients are implemented in the hybrid approach.

\section{Conclusions and Outlook}
\label{sec:conclusions}
In this work a novel modular hybrid approach based on the hadronic transport approach \smsh coupled to the 3+1D viscous hydrodynamics model \vhlle was introduced \cite{hybrid_github}. It can be applied to theoretically model heavy-ion collisions ranging from \sqrts = 4.3 GeV up to \sqrts = 5.02 TeV. In this work, the \hybrid was thoroughly validated regarding consistency at the interfaces, global conservation laws and agreements with other, similar hybrid approaches. Furthermore, the \smsh hadron resonance gas equation of state is constructed and provided as a table, which constitutes a major ingredient for the particlization process. This equation of state, providing the mapping $(e, n_\mathrm{B}, n_\mathrm{Q}) \to (T, p, \mu_\mathrm{B}, \mu_\mathrm{Q}, \mu_\mathrm{S})$, is available under \cite{eos_github}. \\
In continuation, the \hybrid was successfully applied to calculate \dndy and \dndmt spectra at \sqrts = 4.3, 7.7, and 17.3 GeV of $\pi^-$, $p$ and $K^-$. For \dndy as well as \dndmt spectra, the agreement with experimental measurements was improved once employing the \hybrid instead of a pure transport evolution relying on \texttt{SMASH}. In particular, it was demonstrated that the \hybrid is capable of properly capturing the longitudinal baryon dynamics at the aforementioned collision energies.
Furthermore, excitation functions for $\mathrm{d}N/\mathrm{d}y|_{y = 0}$ as well as $\langle p_\mathrm{T} \rangle |_{y = 0}$ were extracted from the \hybrid between \sqrts = 4.3 GeV and 200.0 GeV. Agreement with experimental data improved for $\pi^-$, $p$ and $K^-$ as compared to employing a pure \smsh evolution, especially in view of baryon dynamics.
The integrated $v_2$ and $v_3$ of charged particles as a function of collision energy were further confronted with STAR data for different centralities. A good agreement was found for $v_2$ in central collisions as well as at high collision energies. Otherwise, $v_2$ as well as $v_3$ were underestimated, which can be explained with a too short lifetime of the hydrodynamical fireball as well as with the smearing kernel employed at the initialization of hydrodynamics. \\

The \hybrid was successfully applied to study a broad range of hadronic observables relying on initial conditions extracted on a hypersurface of constant proper time. This assumption is however questionable at low collision energies, where the dynamics tend to be comparably slow. The extension of the \hybrid by more dynamical initial conditions, similar to efforts made in \cite{Akamatsu:2018olk}, thus constitute the next step, to improve the applicability at FAIR/NICA collision energies. In the future, the \hybrid can be applied with different equations of state to evolve the hot and dense fireball - with or without a critical end point or first order phase transition - as well as with varying transport coefficients to systematically study the impact on particle spectra, flow, or other observables.

\begin{acknowledgements}
This work was supported by the Deutsche Forschungsgemeinschaft (DFG, German Research Foundation) – Project number 315477589 – TRR 211 and the DFG SinoGerman project - Project number 410922684.
A.S. acknowledges support by the Stiftung Polytechnische Gesellschaft Frankfurt am Main as well as the GSI F\&E program. A.S. is further grateful to Markus Mayer for providing the flow analysis framework.
I.K. acknowledges support by the Ministry of Education, Youth and Sports of the Czech Republic under grant “International Mobility of Researchers – MSCA IF IV at CTU in Prague” No. CZ.02.2.69/0.0/0.0/20 079/0017983. Computational resources have been provided by the GreenCube at GSI.
This work is further part of a project that has received funding from the European Union’s Horizon 2020 research and innovation program under grant agreement STRONG – 2020 - No 824093 and supported by the State of Hesse within the Research Cluster ELEMENTS (Project ID 500/10.006).
\end{acknowledgements}

\appendix
\section{Centrality Dependence}
\label{app:centrality_dep}
In addition to the results presented in Sec.~\ref{sec:exc_funcs}, the \meanpt excitation functions of $\pi^-$, $p$ and $K^-$ were extracted from the \hybrid for different centralities between 0-5\% and 40-50\%.
\begin{figure}
  \centering
  \includegraphics[width = 0.47\textwidth]{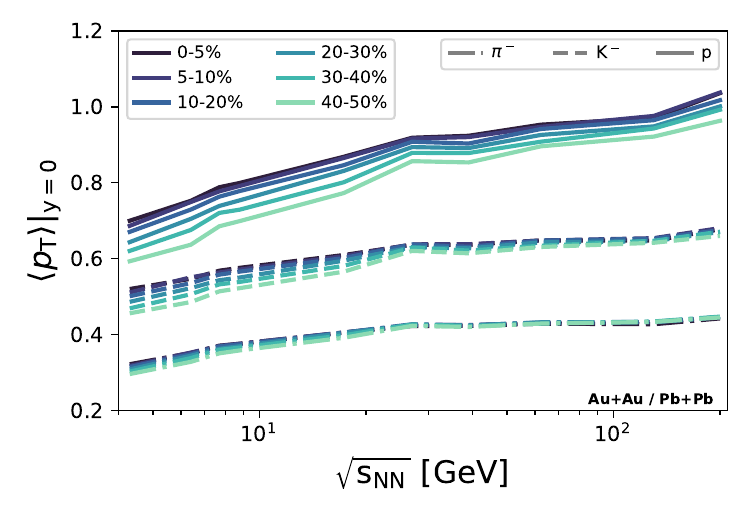}
  \caption{Centrality dependence of mean \pt excitation functions of pions (dashed-dotted), kaons (dashed) and protons (solid). More central collisions are displayed with darker colours, more peripheral collisions with lighter colours.}
  \label{fig:mean_pT_cent}
\end{figure}
These results are presented in Fig.~\ref{fig:mean_pT_cent}, where $\pi^-$ are displayed by dashed-dotted lines, $p$ with solid lines and $K^-$ with dashed lines. Darker colors denote more central collisions, lighter colours more peripheral collisions. It can be observed that the \meanpt of all three particle species generally depend little on the centrality of the collision, yet more peripheral collisions result in smaller mean transverse momenta of pions, protons and kaons. The centrality dependence increases with decreasing collision energies and is most pronounced in the case of protons. \\
The properties of the freezeout hypersurface of the hydrodynamical stage, as presented for central collisions in Fig.~\ref{fig:freezeout_diagram}, can further be analyzed as a function of collision centrality.
\begin{figure}
  \centering
  \includegraphics[width = 0.43\textwidth]{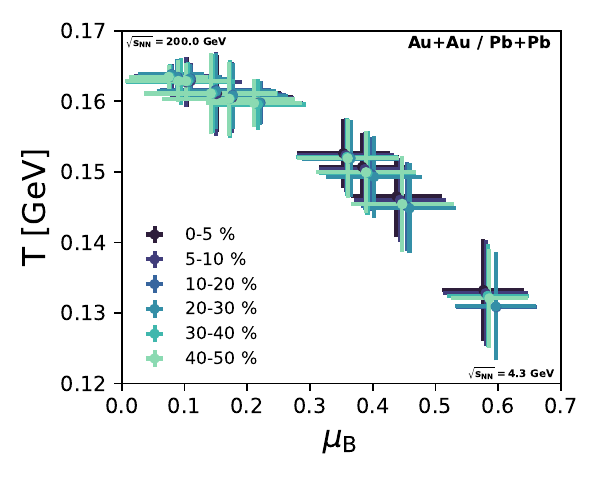}
  \caption{Centrality dependence of the properties of the hydrodynamical freezeout hypersurface. More central collisions are represented by darker colours, more peripheral collisions by lighter colours. Lower collision energies can be found in the lower right corner, higher collision energies in the upper left corner.}
  \label{fig:freezeout_diagram_cent}
\end{figure}
This is displayed in Fig.~\ref{fig:freezeout_diagram_cent}, where central Au+Au/Pb+Pb collisions are again represented by darker coloured markers, more peripheral collisions by lighter coloured markers. The collision energies rise from \sqrts = 4.3 GeV in the lower right corner to \sqrts = 200.0 GeV in the upper left corner.
It can be observed that the general dependence on the collision centrality is moderate. Especially for high-energy collisions, the freeze-out properties are nearly unaffected by centrality variation. Towards lower collision energies a stronger centrality dependence is observed, but it is small.

\section{Parametrizing the \smsh hadron resonance gas equation of state}
\label{app:parametrization}
As mentioned in Sec.~\ref{sec:SMASH-EoS} we have made different attempts to parametrize the \smsh equation of state for it to be applied more easily within the \hybrid and potentially also in other works.
Unfortunately, these efforts remain unsuccessful up to now. We still want to briefly document the different attempts we made in the following. \\
The purpose of the equation of state is to perform the mapping $(e, n_B, n_Q)\rightarrow(T, p, \mu_B, \mu_Q, \mu_S)$.
Currently, the coupled equations (c.f. Eqs.~\ref{eq:EoS_equations}) are solved once for different combinations of $e$, $n_\mathrm{B}$ and $n_\mathrm{Q}$ and stored in a table, where the resulting thermodynamic quantities need to be looked up for a given set of $(e, n_\mathrm{B}, n_\mathrm{Q})$ and interpolated between the grid points, if necessary.
This procedure is time-consuming and introduces uncertainties related to finite grid effects when interpolating between the predefined grid points. An alternative solution would be to directly solve the coupled thermodynamic equations directly within the hydrodynamical simulation for each cell individually, relying on its specific densities. This procedure is however even less effective and requires large computational resources. A parametrization from which one could directly calculate the thermodynamic quantities and that relies on a limited number of parameters only would thus be the most effective solution. \\
We have tried to use the following polynomial-inspired ansatzes in order to find a parametrization of the \smsh hadron resonance gas equation of state: \\
\begin{itemize}
  \item[1.] $f(e, n_B, n_Q) = \sum_{i, j, k = 0}^{N_{\rm{Dim}}} a_{ijk} e^i n_B^j n_Q^k$
  \item[2.] $f(e, n_B, n_Q) = \tfrac{\sum_{i, j, k = 0}^{N_{\rm{Dim}}} a_{ijk} e^i n_B^j n_Q^k}{\sum_{i, j, k = 0}^{N_{\mathrm{Dim}}} b_{ijk} e^i n_B^j n_Q^k} + c_0$
  \item[3.] $f(e, n_B, n_Q) = \sum_{i, j, k = 0}^{N_{\rm{Dim}}} \ a_{ijk} \ \mathrm{log}(b_{ijk} \ e^i) \ n_B^j \ n_Q^k$ \\[-0.3cm]
  \item[4.] $f(e, n_B, n_Q) = \sum_{i, j, k = 0}^{N_{\rm{Dim}}} \ a_{ijk} \ (b_{ijk} \ e^i)^{-c_{ijk}} \ n_B^j \ n_Q^k$ \\
\end{itemize}
Here, $a_{ijk}$, $b_{ijk}$, or $c_{ijk}$ denote the sets of parameters we want to determine. The major problem we faced was the rapidly growing number of parameters $N_{\rm{Param}}$ for larger $N_{\rm{Dim}}$ (for example in the second case $N_{\rm{Dim}} = 3$ results in $N_{\rm{Param}} = 55$). Large values for $N_{\rm{Dim}}$ are however necessary as the equation of state is a non-trivial function and small values for $N_{\rm{Dim}}$ resulted in unsatisfactory parametrizations.
In addition, a good initial approximation of $a_{ijk}$, $b_{ijk}$, and $c_{ijk}$ is essential for the solver to succeed. Besides a uniform sampling in parameter space, we have also explored a Latin Hypercube Sampling (LHS) procedure \cite{10.2307/1268522}, where the parameter space is covered in a more efficient way to provide better initial guesses. Both ways of determining initial guesses and later checking for a best fit did however not result in any satisfying parametrization of the underlying equation of state.
On top of this we utilized Chebyshev nodes as grid points to minimize strong oscillations at the edges of the equation of state, which did not alleviate our issues either. \\
In summary, wie explored different possibilities to find a suitable parametrization for the \smsh hadron resonance gas equation of state but did unfortunately not succeed. We thus fall back to our look-up table that requires interpolations between the grid points. This table is available under \cite{eos_github}.



\end{document}